\documentclass[vecphys]{svmult}		

\usepackage{makeidx}         
\usepackage{graphicx}        
\usepackage{epsfig}          
\usepackage{multicol}        
\usepackage[bottom]{footmisc}
\usepackage{amsmath}
\usepackage{subfigure}
\usepackage{color}
\usepackage{multicol}
\usepackage{caption}

\DeclareCaptionLabelFormat{andtable}{#1~#2  \&  \tablename~\thetable}

\newcommand{\sech}{\text{sech}}

\makeindex             
\begin{document}

\def\jcmindex#1{\index{#1}}
\def\myidxeffect#1{{\bf\large #1}}

\title*{$\phi^4$ Solitary Waves in a Parabolic Potential: Existence, Stability, and Collisional Dynamics}
\titlerunning{$\phi^4$ Solitary Waves With a Parabolic Potential}
\author{
R.M. Ross\inst{1}
\and
P.G. Kevrekidis\inst{1}
\and
D.K. Campbell\inst{2}
\and
R.Decker\inst{3}
\and
A. Demirkaya\inst{3}
}

\institute{
Department of Mathematics and Statistics, University of Massachusetts,
Amherst, MA 01003-4515, USA, \texttt{ryanross1023@gmail.com, kevrekid@math.umass.edu}
\and
Department of Physics, Boston University,
590 Commonwealth Ave, Boston, MA, USA, \texttt{dkcampbe@bu.edu}
\and
Mathematics Department, University of Hartford, 200 Bloomfield Ave, West Hartford, CT 06117, USA, \texttt{rdecker@hartford.edu, demirkaya@hartford.edu}
}
\maketitle
%
\abstract
{
We explore a $\phi^4$ model with an added external parabolic
  potential term. This term dramatically alters the spectral properties
  of the system.
  We identify single and multiple kink solutions and examine their
 stability features; importantly, all of the stationary structures
 turn out to be unstable.
 We complement these with a dynamical study of the evolution
 of a single kink in the trap, as well as of
  the scattering of kink and anti-kink solutions of the model. We see that some of the key characteristics of kink-antikink collisions, such as the critical velocity and the multi-bounce windows, are sensitively dependent on the trap strength
  parameter, as well as
  the initial displacement of the kink and antikink. 
}


\section{Introduction}
\label{sec:intro}

Models of the sine-Gordon (sG)~\cite{ussg} and more generally of the nonlinear
Klein-Gordon type, such as $\phi^4$~\cite{belova}
have been studied  intensely via a combination of analytical and numerical
techniques for
well over three decades now. Part of the reason for this widespread appeal
concerns the diverse set of physical applications for which such models have been
argued to be of relevance. These start from the simplest (coupled
torsion pendula, motion of dislocations etc. in the case of the discrete
sG~\cite{braun}) and extend far and wide. Case examples include,
but are not limited to structural phase transitions \cite{behera,gufan},
domain walls arising in cosmological models~\cite{vilenkin01,anninos},
simple polymeric models, as well as models of uniaxial
ferroelectrics; see, e.g., Ch.~9 in~\cite{Vach}.

However, there is another source of the fascination, especially from an
applied mathematics perspective, in
non-integrable models such as $\phi^4$. This has to do
with their remarkable properties related to the
collisional dynamics of kinks and anti-kinks. There,
resonant phenomena between translational, internal and extended
modes have been identified and and a phenomenological
``resonance energy exchange mechanism"~\cite{campbell}
has been used to characterize
features such as the ``multi-bounce windows", the fractal emergence
of such windows, and the universal nature of these phenomena
not only in $\phi^4$, but also in other models such as the
double, and the parametric sG equations~\cite{campbell,Campbell1,anninos}.
While recent years have seen both a more rigorous mathematical
analysis~\cite{goodman2} and an experimental realization
of a system reminiscent of the effective kink-antikink
interactions~\cite{goodman}, there have also been further studies causing
a renewed interest in the phenomenological approach and towards
posing new questions.
For instance, particular variations of the $\phi^6$ model
have been  shown numerically to possess multi-bounce windows
without possessing internal modes~\cite{shnir1}.
It has also been found that by parametrically deforming
the $\phi^4$ model, one can introduce more internal
modes and suppress the two-bounce windows~\cite{simas}.
Lastly, the recent work of~\cite{weigel,weigel2}
has performed a careful bookkeeping of the collective coordinate
approaches intended to provide an effective description of the collision
events. In so doing, a number of nontrivial problems regarding
both computing with such reduced models and the conclusions
drawn from them have arisen. Thus, despite its time-honored tradition,
the study of kink-antikink collisions in non-integrable models remains a  surprisingly active field of investigation to
this day~\cite{gani1,danial,gani2,gani3,gani4}.

In the present work, we examine a different variant of the
$\phi^4$ problem, considering the scenario where an external
potential has been imposed on the field theory. In particular,
motivated by the extensive work on atomic Bose-Einstein condensates (BEC),
we consider a parabolic trap, which in the latter context
emulates the role of a generic magnetic
confinement~\cite{pethick,stringari,darkbook}.
This possibility turns out to alter dramatically the properties
of the kink waveforms. In particular, even the single kink
turns out to be dynamically unstable in the emerging energy
landscape. We examine the case of the single kink and
its existence, stability and dynamic properties, as well as
that of the kink-antikink state. In the latter, we examine
the kink-antikink collisions and attempt to capture them
by means of low dimensional collective coordinate models.
We highlight both the favorable, as well as the unfavorable
traits of these models with respect to their ability to capture the full PDE dynamics
and raise some relevant possibilities towards future studies.

\section{The Model, Calculational Approach and Ground State}
Our starting point for the considerations that follow is
the classical $\phi^4$ equation with an added $x$-dependent trapping term: 
\begin{align}
	u_{tt} = u_{xx}-{\frac{\partial V}{\partial u}}
	\label{fullpde}
	\end{align}
	where the potential function is given by
	\begin{equation}
	{V(u,x)}=\frac{1}{2}(u^2-1)^2-\frac{1}{2}+\frac{1}{4}\Omega^2 x^2 u^2.
	\end{equation}	
$\Omega$ is the trap strength constant and typically (in settings
related to our motivating example of atomic
BECs~\cite{pethick,stringari,darkbook}) assumes values in $(0,1)$. However, we mostly restrict our attention to the dynamics for $\Omega\ll 1$, using $\Omega=0.15$ for most of the data runs~\footnote{Although the BEC example
is simply of motivating nature, it turns out that in that setting
$\Omega$ plays the role of longitudinal to transverse trapping, hence
to achieve effective one-dimensionality of the system the condition
$\Omega \ll 1$ needs to be enforced. This is what motivates our
selection here too.}. The case $\Omega= 0$ has been well-studied in the past and in its analysis, the perturbation theory and collective coordinate methods have been used \cite{anninos,belova,campbell,sugiyama,weigel}.  We extend these methods to study our $\phi^4$ variant bearing a trapping in different parameter regimes.

To explore the properties of Eq. \eqref{fullpde} we employ the following 
numerical scheme. First, we use a fourth-order central difference in space (see e.g. \cite{anninos,roy} for this and other similar difference schemes).
The resulting system of ordinary differential equations is then evolved
in time using fourth-order Runge-Kutta methods, ensuring the
accuracy of the conservation of the energy.
 {The average value of the energy is of O(1), while the deviations from the mean are no more than  O($10^{-4}$). }
We apply  free boundary conditions, although as we shall see below this is generally
inconsequential because the field effectively
vanishes well before the edge of our computational domain.
For all our data runs, a spatial step size of $\Delta x = 0.02$ was used.
Our computations typically use  $\Omega = 0.15$
and a spatial grid running in the interval $[-30, 30]$.
For smaller values of $\Omega$, we use a larger interval as necessary
 to ensure the decay of the wavefunction within the computational
domain.

The stationary ground state of the system $u_{\Omega}(x)$ solves the
ordinary differential equation (ODE)
\begin{align}
u_{xx} + 2(u-u^3) - \frac{1}{2}\Omega^2 x^2 u = 0
\label{eq:utt=0}
\end{align} 

$u_{\Omega}$ is not known explicitly but in the case $\Omega \ll 1$, it is well-estimated by the Thomas-Fermi (TF) approximation which neglects
the spatial variation of $u$ (again
motivated by similar considerations in atomic
condensates~\cite{pethick,stringari}) defined by

\begin{align}
u_{\text{TF}}(x) = \max\bigg\{0, \sqrt{1-\frac{1}{4}\Omega^2 x^2} \bigg\}
\label{eq:TFapprox}
\end{align} 

\begin{figure}
	\centering
		\includegraphics[width=\textwidth]{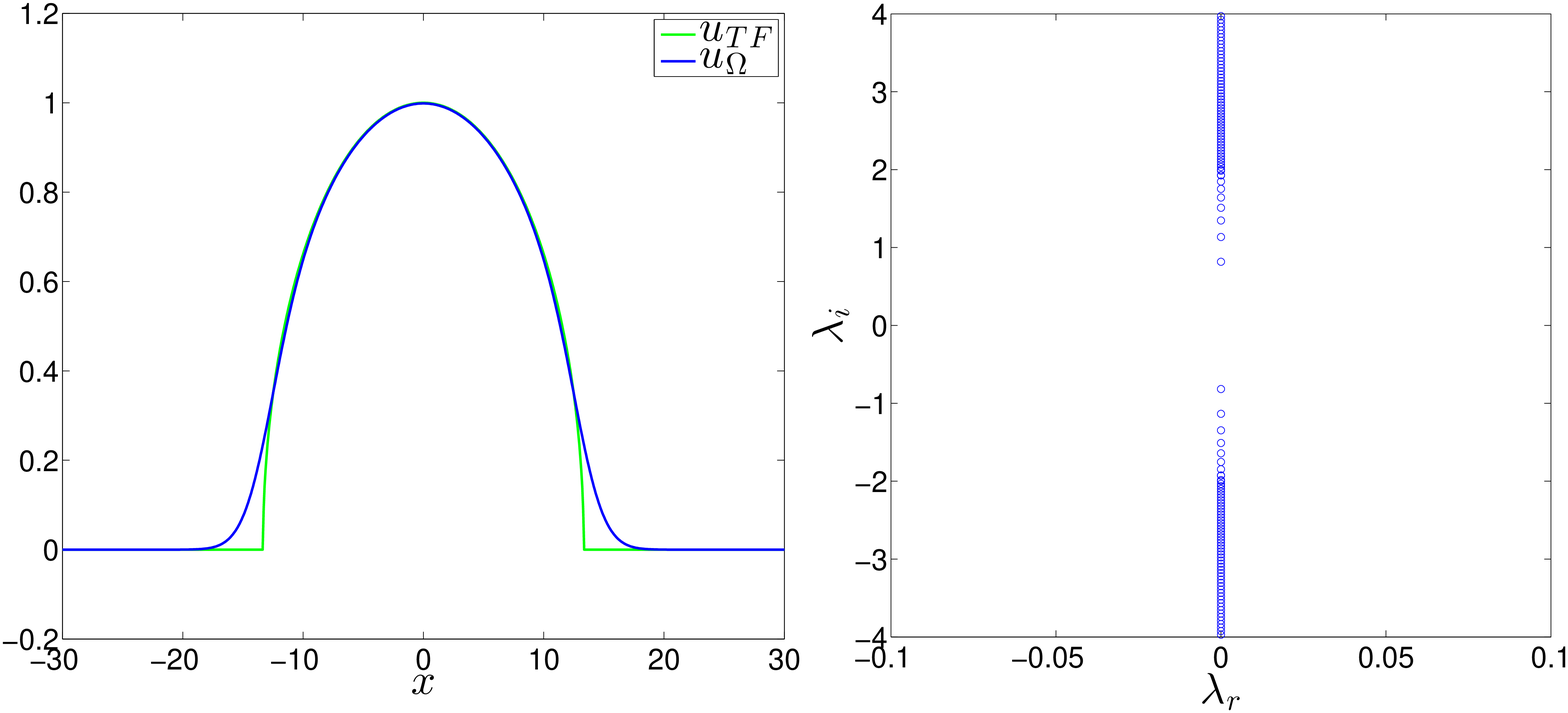}
	\caption{ The left panel shows the Thomas-Fermi approximation $u_{\text{TF}}(x)$ (in green) from \eqref{eq:TFapprox} and the
        numerical
        ground state $u_{\Omega}$ obtained through the Newton-Raphson
        iterations (in blue) for $\Omega=0.15$. The right panel shows the
        spectral plane $(\lambda_r,\lambda_i)$ of linearization
        eigenvalues $\lambda=\lambda_r+i \lambda_i$
        corresponding to the ground state; the eigenvalues are all purely imaginary, indicating the spectral stability of the ground state.} 
	\label{u_omega}
\end{figure}

Eq.~(\ref{eq:TFapprox}) solves Eq.~\eqref{eq:utt=0} when the $u_{xx}$ term is neglected. We use this approximation as our initial guess for the Newton-Raphson
iterations converging to the true numerical ground state of the system.
The comparison of the TF approximation with the true numerical result is
given in Fig.~\ref{u_omega}.
At the level of the TF approximation, we note that the ground state is zero whenever $|x|>x_s$, where $u_{\text{TF}}(x_s)=0$, which gives
\begin{align}
x_s = \frac{2}{\Omega}
\label{eq:x_supp}
\end{align}
As we can observe in the figure, the TF approximation is very good
except for the vicinity of $x_s$, where its disagreement with the
true ground state has been quantified in a series of rigorous
studies~\cite{gallo,karali}.

To study the stability of the ground state, we consider the linearization around
the solution using the ansatz:
\begin{align}
u(x,t) = u_{\Omega}(x) + \epsilon e^{\lambda t} \chi(x).
\label{ansatz_par}
\end{align}
Here $u_{\Omega}$ is the numerically obtained ground state, and $(\lambda,\chi)$
represent
the corresponding eigenvalues and eigenvectors of the linearization.
The equation that these satisfy at $O(\epsilon)$ is of the form:
\begin{align}
\lambda^2 \chi = \chi'' -V''(u_{\Omega})\chi \label{eq:lin}
\end{align}
For the ground state we find, as is shown in the right panel of Fig.~(\ref{u_omega})
that all the eigenvalues lie on the imaginary axis, hence the relevant
state is spectrally stable; for instability to arise, at least one
of the relevant eigenstates should correspond to growth, i.e.,
to Re$(\lambda)>0$.

\section{Single Kink Solutions}
\begin{figure}[!ht]
 \begin{center}
	\includegraphics[scale=0.2]{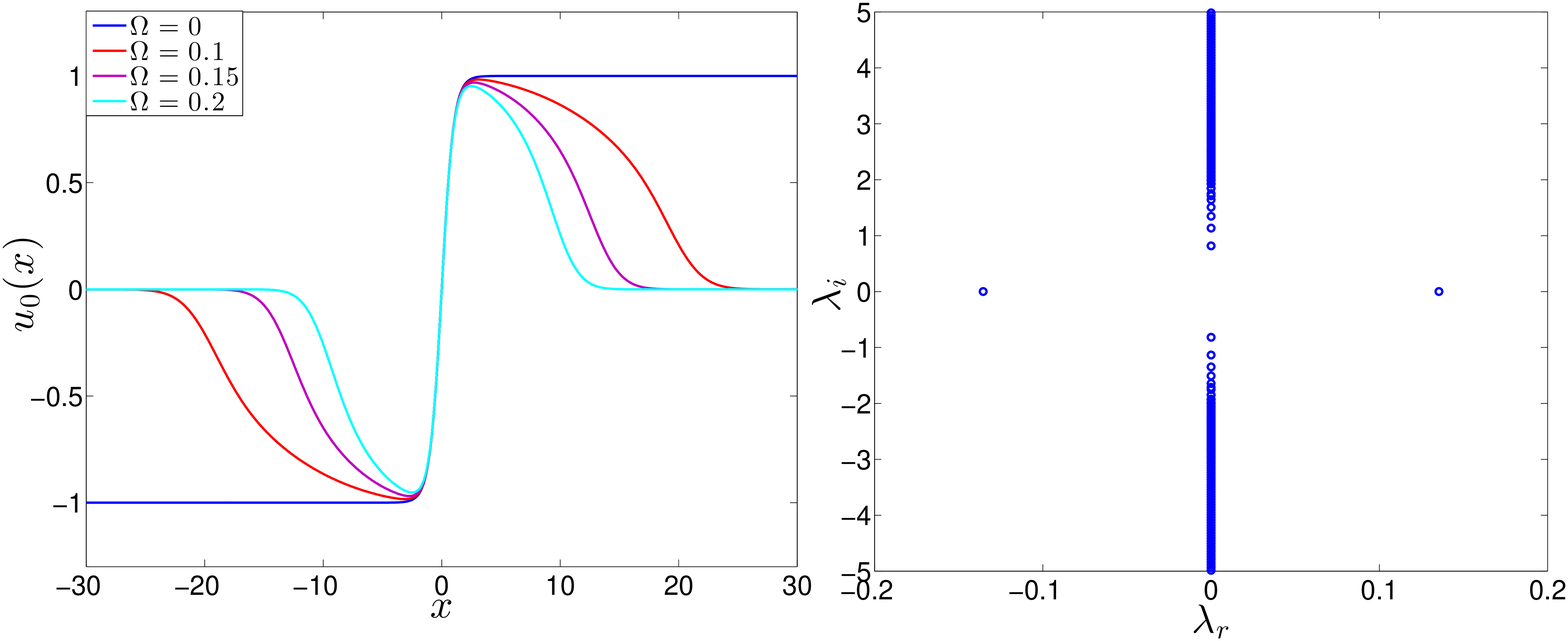}
	\caption{ The left panel shows the single kink solution as $\Omega$ varies. The right panel shows the spectral plane when $\Omega=0.15$. The nonzero
        real eigenvalues are $\approx \pm 0.13$. }
	\label{singkink}
	\end{center}
\end{figure}

We now turn to excited states in the system in the form of kinks
and anti-kinks, as well as bound states thereof.
We start with the single static kink whose solution is
not known explicitly but can be numerically approximated by 
\begin{equation}u_{0,K} (x) \approx  u_{\Omega}(x) \tanh (x), \hspace{0.5cm} u_{0,\tilde{K}} (x) \approx  -u_{\Omega}(x) \tanh (x),
\label{static_kink}
\end{equation}
once again in the TF approximation.
For various values of $\Omega$, the numerically calculated solutions are shown in Figure \ref{singkink}.

In the figure, also the spectrum of linearization around the single kink
is shown (right panel). We see that in all the cases of
 nonzero $\Omega$, there exists a pair of nonzero real eigenvalues.
More concretely, the eigenvalue pair that used to be at $\lambda=0$
due to translational invariance now exists as real and leads to the
instability of the kink, which is progressively stronger,
the larger the value of $\Omega$.

\subsection{Numerical Results (PDE)}

We now examine the dynamical evolution of such unstable kinks.
As a means of obtaining moving kink solutions $u_{K}$ (or antikink solutions $u_{\tilde{K}}$) of (\ref{fullpde}), we can employ the
Lorentz transformation, $x\to  \gamma(x-x_0-vt)$
where $\gamma=1/\sqrt{1-v^2}$, where
$x_0$ is the initial position and $v$ is the velocity of the moving kink. The initial conditions of the PDE are
\begin{align*}u(x,0)&=u_{\Omega}(x) \tanh(\gamma (x-x_0)); \\
u_t(x,0)&=-v\gamma u_{\Omega}(x) \sech^2(\gamma(x-x_0));
\end{align*}
Of course the model in the presence of the potential
is no longer Lorentz invariant (as it is in the case of $\Omega=0$).
Nevertheless, our results have shown that
from a numerical perspective producing a Lorentz-boosted
kink and multiplying it by the (stationary) TF background is
an efficient method to produce ``moving'' waveforms
starting from standing ones.

\begin{figure}[tbp]
\begin{center}
      \subfigure[]{{\includegraphics[width=0.49\textwidth]{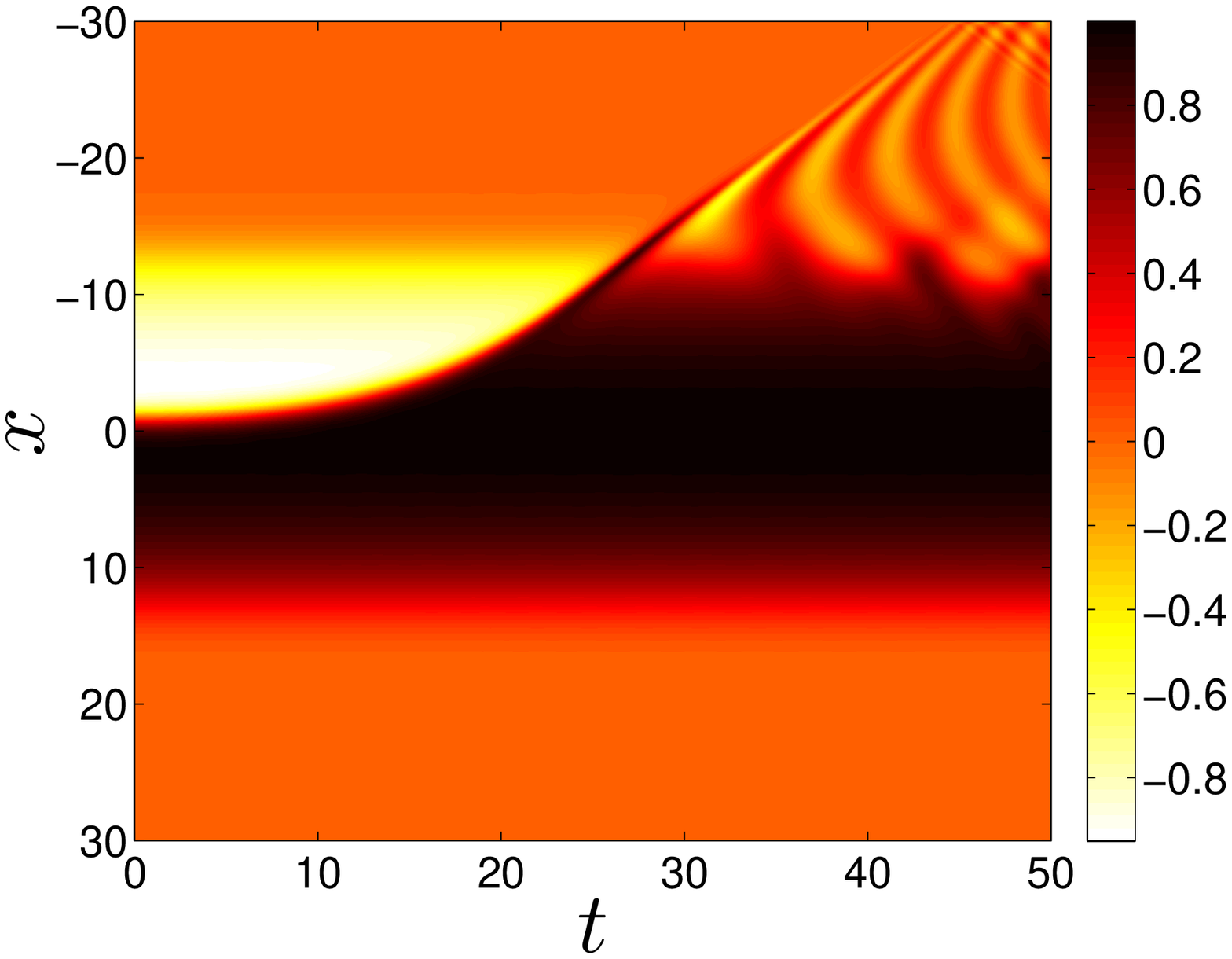}}}
      \subfigure[]{{\includegraphics[width=0.48\textwidth]{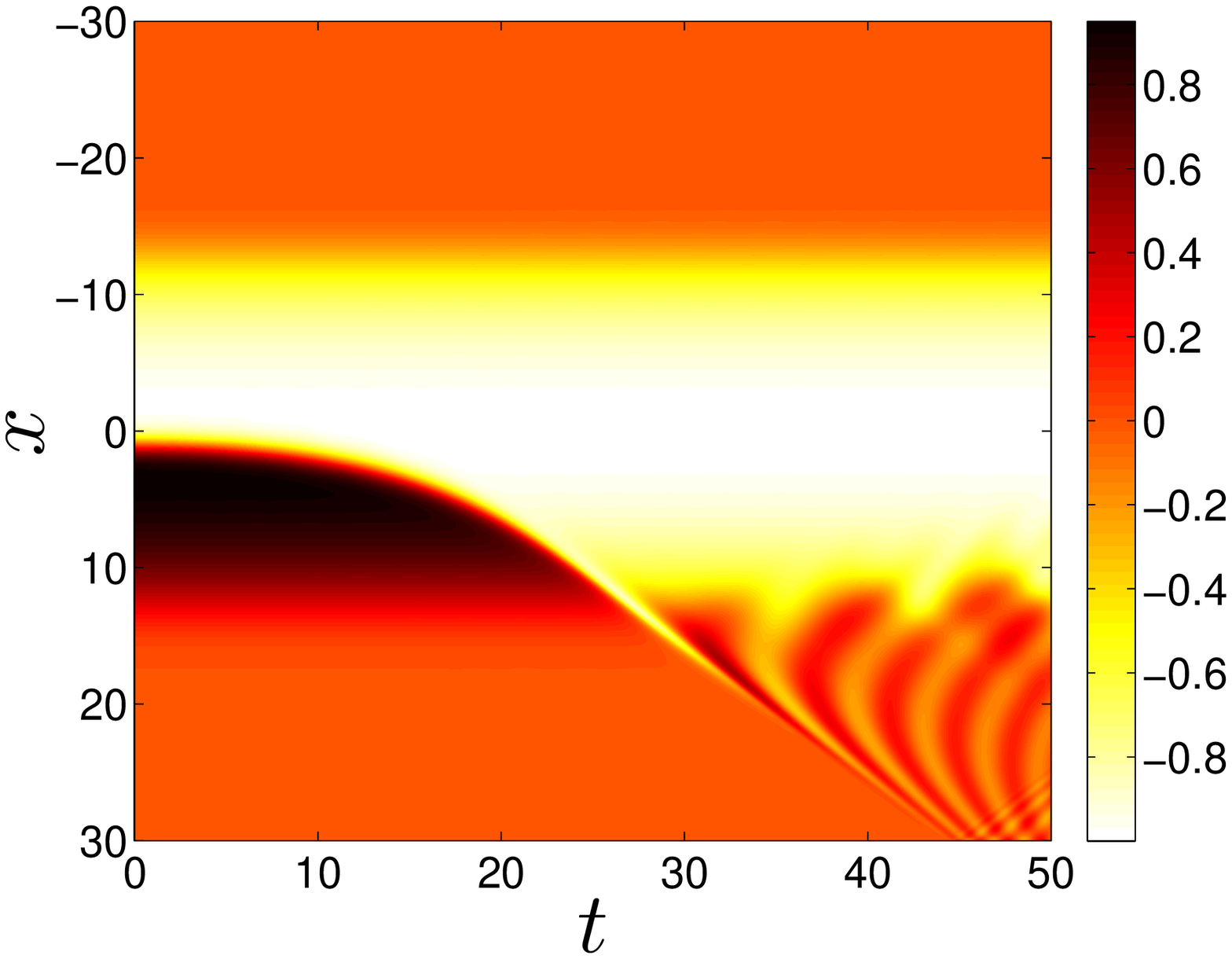}}}
       \subfigure[]{{\includegraphics[width=0.49\textwidth]{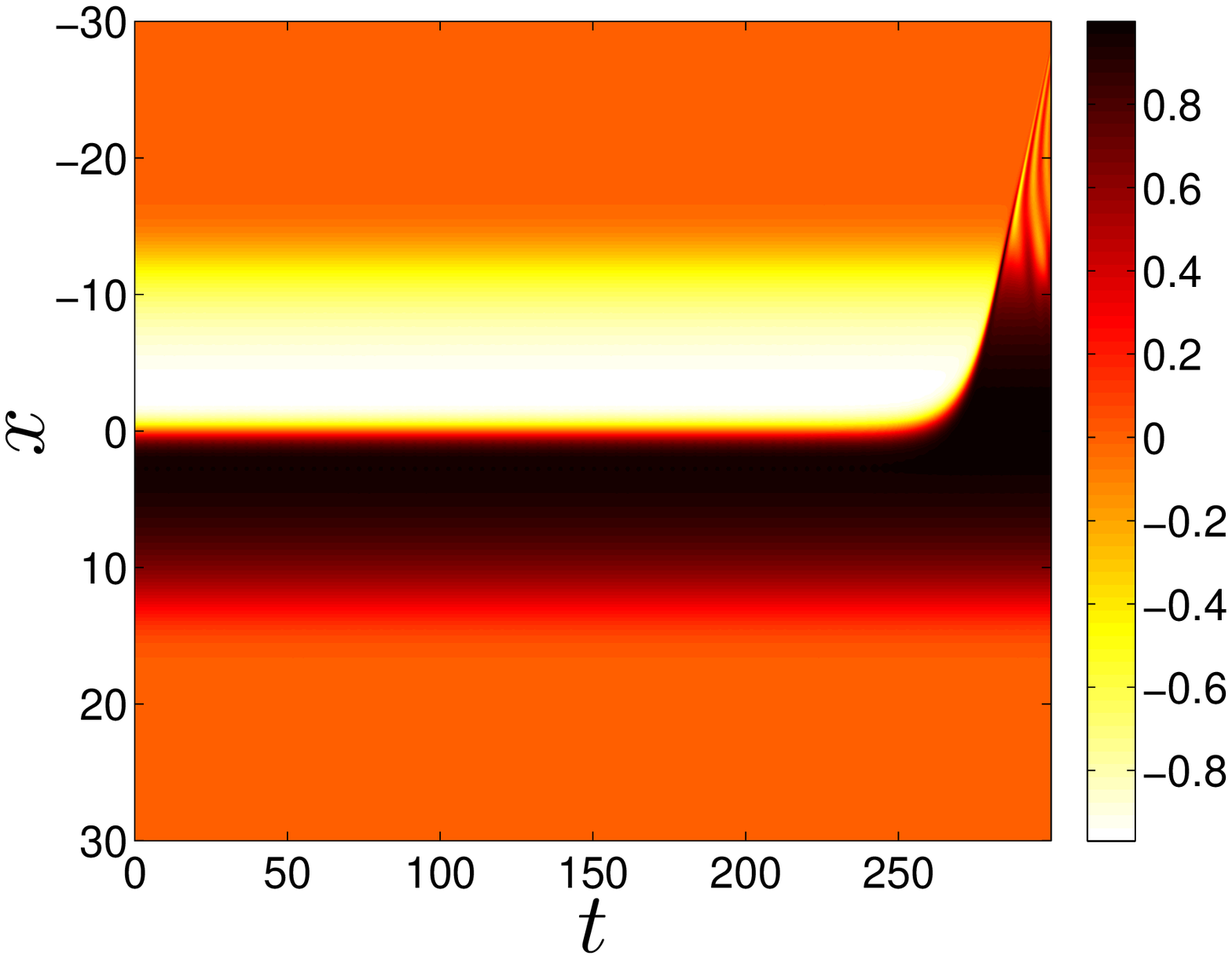}}}
  \end{center}
     \caption{(a) Initialized with center position
       $x_0=-1$, the kink slides away from zero -- to the left, because $x_0$ is negative -- and gradually accelerates. (b)  With $x_0=1$, the kink slides away from zero to the right, because $x_0$ is positive and gradually accelerates. (c) With $x_0=0$, the kink preserves its steady form for a while but eventually
       as its instability manifests,
    it slides (in this particular case) to the right.}
    \label{right_left}
\end{figure}

The presence of the the trapping term causes the velocity of the kink not to remain constant
(as it would under stationary initialization in the untrapped, translationally
invariant problem). When the initial velocity ($v_\mathrm{in}$) is zero,
the numerical results show that the motion of the kink solution depends sensitively on
its starting position. When the kink starts to the left of the origin
($x_0>0$), the kink slides to the left, accelerating as it does so. If the kink starts to the right of the origin ($x_0<0$), then it slides to the right. In both cases, the kink eventually ends up being expelled from the system.
Fig.~\ref{right_left} shows examples of initializing the kink both
to the left and to the right of the fixed point {at $x=0$ } with no speed and
illustrates how the kink slides along the side where it starts.
This corroborates in a definitive way the results of the stability
analysis in that the kink now (in the presence of the parabolic
trap) encounters a saddle point at $x=0$ (the center of the trap).
Hence, on each side of this unstable equilibrium, it will slide
along the unstable manifold. In fact, it is interesting to note
that in panel (c) of the figure the kink is initialized {\it at}
$x=0$ (up to numerical roundoff error). We can see that in that
case, it stays at the origin for a long time, until eventually
the projection of the roundoff error in the unstable eigendirection
grows to O$(1)$ and kicks the kink away from the unstable equilibrium
(in the case of this particular example its center moves to $x>0$,
but it can just as well move to the $x<0$, depending on the perturbation).

\begin{figure}[tbp]
\begin{center}
      \subfigure[]{{\includegraphics[width=0.49\textwidth]{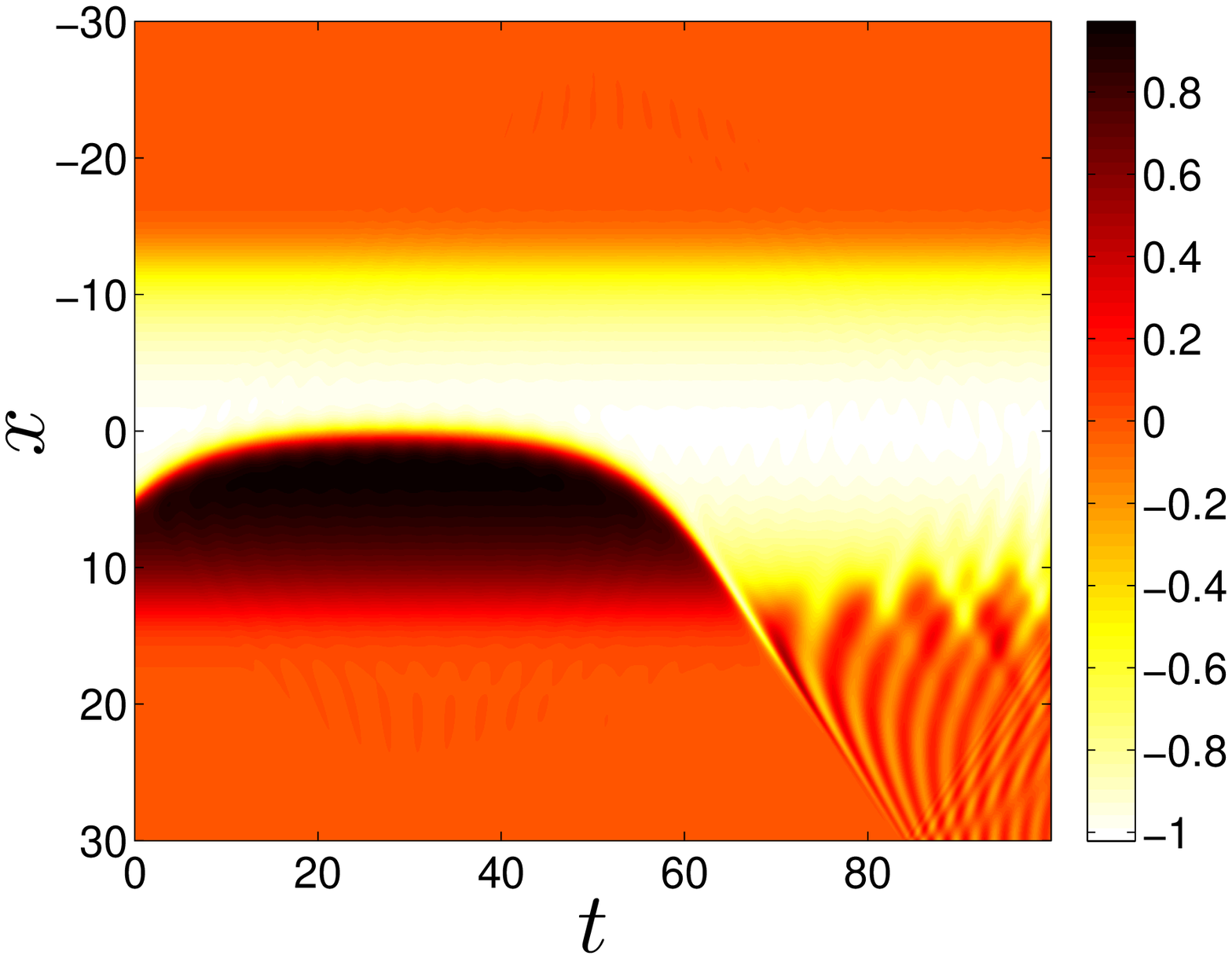}}}
      \subfigure[]{{\includegraphics[width=0.49\textwidth]{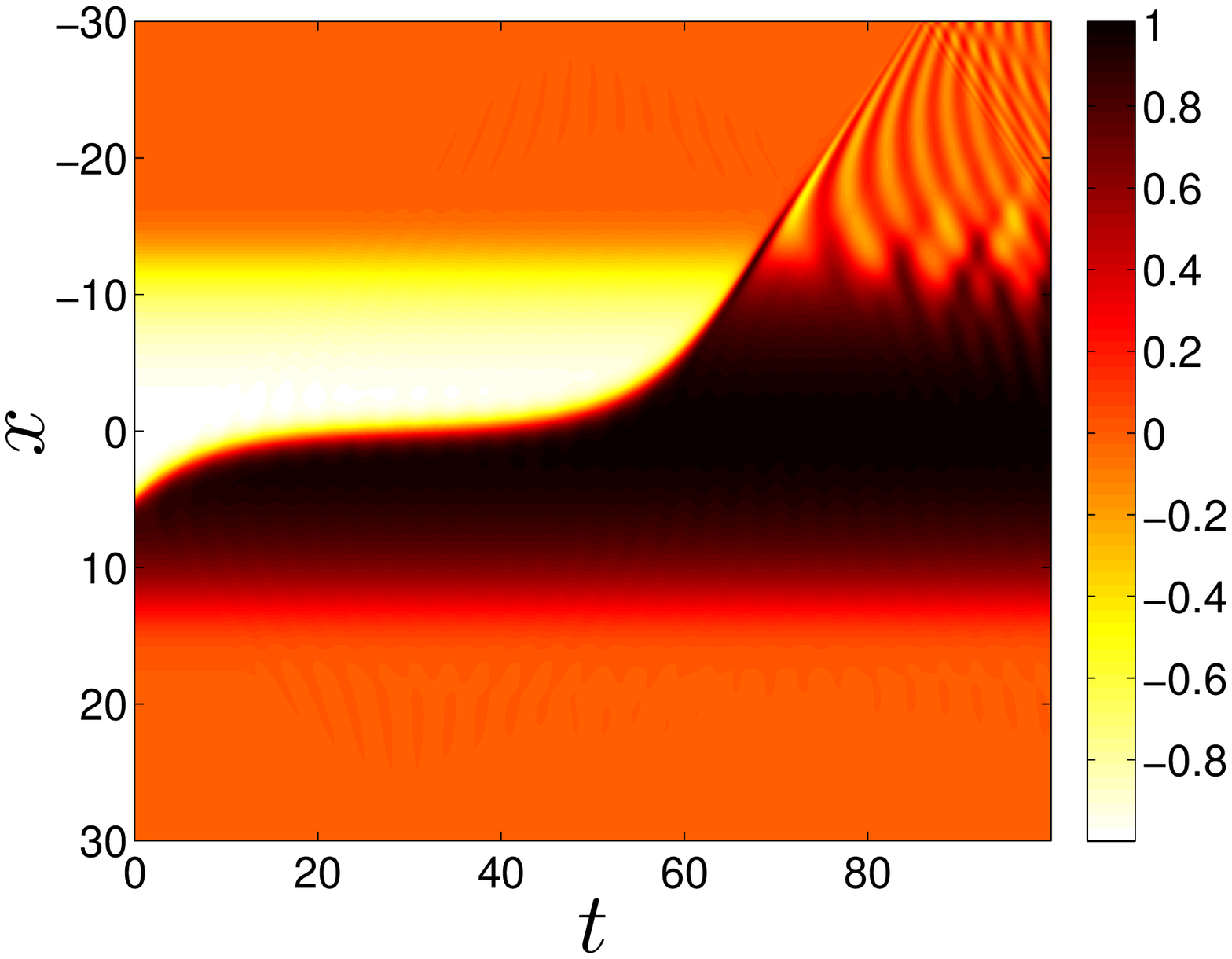}}}
      \subfigure[]{{\includegraphics[width=0.49\textwidth]{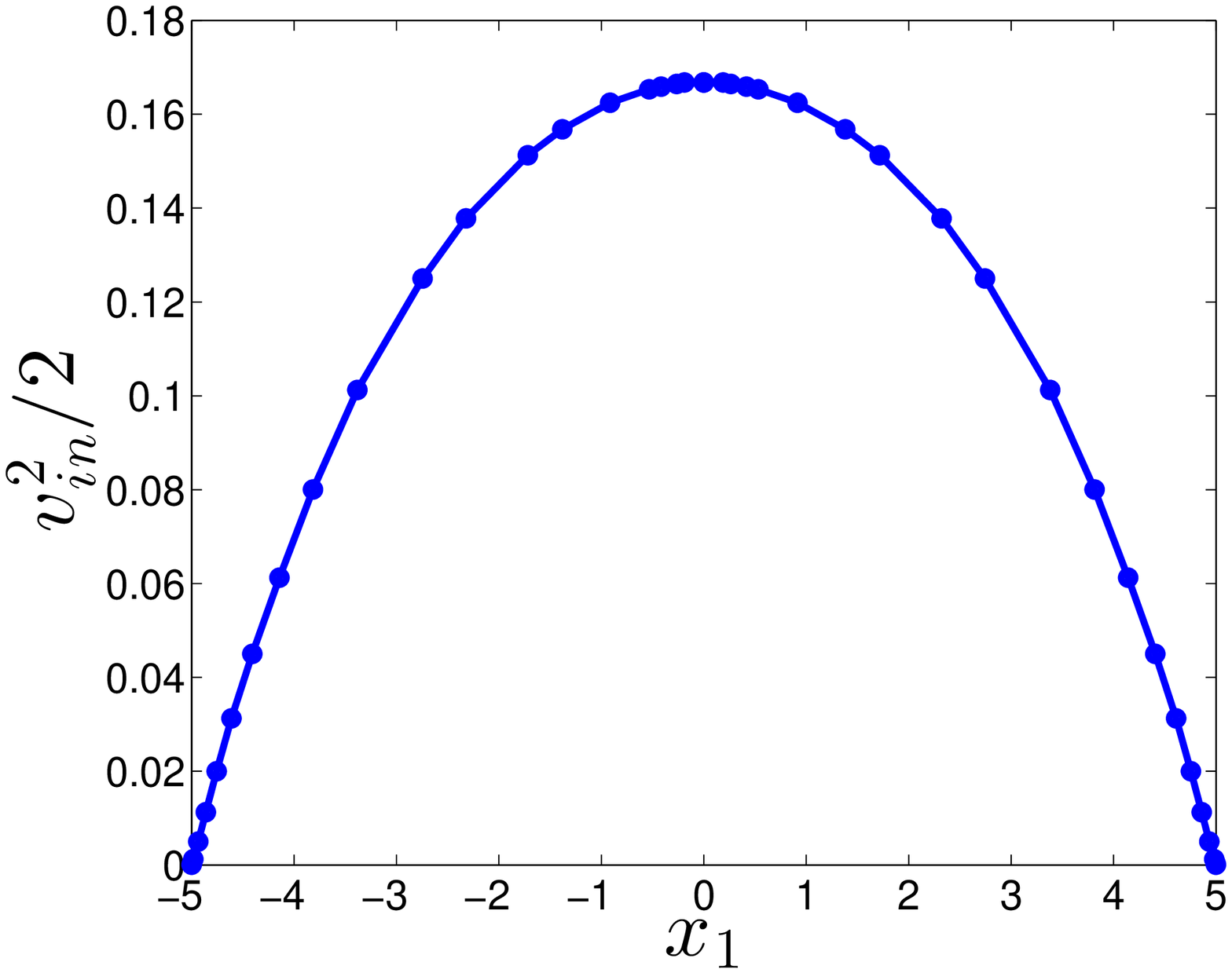}}}

  \end{center}
   \caption{ A moving kink with $x_0=5$ initially with (a) $v_{in}=0.577$, lacks the kinetic energy to overcome the potential energy
    barrier at $x=0$. On the other hand, in (b) with
    $v_{in}=0.578$, its kinetic energy suffices for
    it to break through to the other side. {(c) shows the effective potential of the moving kink with $x_0=5$, approximated as explained in the text. } }
    \label{nonzerovelocity}
\end{figure}

With nonzero initial velocity, the behavior of the kink depends on the magnitude of the velocity. For a kink starting away from the origin, the saddle point
at the $x_0=0$ represents a finite (potential) energy barrier which the kink must overcome to break through to the other side.
Given sufficient kinetic energy, a kink moving towards the origin will
overcome this potential energy barrier.
On the other hand, if the velocity (and hence the kinetic energy of the kink is  ``subcritical", i.e., below the value needed to overcome the potential
barrier) then the kink will bounce off the barrier and return
to be expelled on the side that it started. Fig. \ref{nonzerovelocity} (a),(b)
demonstrate these two scenarios in a convincing way numerically.
In particular, between $v_\mathrm{in}=0.577$ and $v_\mathrm{in}=0.578$
of the left and right panel, respectively, there is clearly a critical
velocity as the former results in a bounce-back of the kink from $x=0$,
while the latter leads to its transmission through the barrier and on
to the other side.  {For any $v_\mathrm{in}<0.577$, the kink stops at $x=x_1$, called a turning point and then moves in the opposite direction. The Fig. \ref{nonzerovelocity} (c) shows the relation between the effective potential
  approximated by the associated kinetic energy
  (per unit mass) $\frac{1}{2}v_\mathrm{in}^2$ and the turning point $x_1$.  }
Of course, in a way consistent with the above picture,
if the kink is initially moving away from the origin, then it will continue
to move away and gradually accelerate over time.

\subsection{Collective Coordinate Approach (ODE)}

In this section we formulate a ``collective coordinate" (CC) approach for the
dynamics for the single kink. Our aim is to reduce the full PDE with
infinitely many degrees of freedom to 
an ODE
model with a single degree of freedom. This is partly driven by
previous CC studies of the motion of a kink
in the homogeneous case ($\Omega = 0$), 
{including the work of Sugiyama \cite{sugiyama}
(see also~\cite{anninos})}. It is also importantly motivated
by the
nature of the motion in the vicinity of the
saddle point discussed above, and by the desire to obtain a simplified description
of this motion.

To begin, we note that the full PDE conserves the energy (Hamiltonian)
\begin{align}
\mathcal H = \frac{1}{2}\int_{-\infty}^\infty u_t^2 + u_x^2 + (u^2-1)^2 + \frac{1}{2}\Omega^2 x^2u^2 \, dx = \mathcal T + \mathcal V,
\label{eq:H_par}
\end{align}
where the kinetic and potential energies of the kink, respectively, are
\begin{align*}
&\mathcal{T}(t) = \frac{1}{2} \int_{-\infty}^\infty u_t^2 \, dx 
\\
&\mathcal{V}(t) = \frac{1}{2} \int_{-\infty}^\infty u_x^2 + (u^2-1)^2 + \frac{1}{2}\Omega^2 x^2u^2 \, dx. 
\end{align*}
 
The corresponding Lagrangian is
\begin{equation}\label{lagr}
\begin{aligned}
\mathcal{L}(t) &= \mathcal {T}(t) - \mathcal {V}(t)\\
&= \frac{1}{2}\int_{-\infty}^\infty u_t^2 - u_x^2 - (u^2-1)^2 - \frac{1}{2}\Omega^2 x^2u^2 \, dx.
\end{aligned}
\end{equation}

Assuming that we operate in the TF limit of $\Omega \ll 1$,
we seek to characterize the kink motion by utilizing  the ansatz
\begin{align*}
u_1(x,t) = u_\Omega(x) \tanh(x - X(t)),
\end{align*}
where $X(t)$ is the time-dependent displacement of the kink from the origin and $u_{\Omega}$ is the ground state of the system in the presence of the trap.
Then, the Lagrangian $\mathcal L$  becomes as follows:
\begin{align*}
\mathcal{L}(X,\dot X) = \int_{R} L(u_1(x,X(t))) \, dx =  a_0(X)\dot X^2 - a_1(X), 
\end{align*}

where
\begin{align*}
&a_0(X) = \frac{1}{2}\int_{R} u_{\Omega}^2(x)\sech^4(x-X) \, dx \\
&a_1(X) = \frac{1}{2} [I_1(X) + I_2(X) + I_3(X)+I_4(X)+I_5(X)] \\
&I_1(X) = \int_{R} [u_{\Omega}'(x)]^2 \tanh^2(x-X) \, dx \\
&I_2(X) = \int_{R} 2u_\Omega'(x)u_\Omega(x)\tanh(x-X)\sech^2(x-X) \, dx \\
&I_3(X) = \int_{R} u_{\Omega}^2(x)\sech^4(x-X) \, dx \\
&I_4(X) = \int_{R} (u_{\Omega}^2(x)\tanh^2(x-X)-1)^2-1 \, dx \\
&I_5(X) = \int_{R} \frac{1}{2}\Omega^2 x^2 u_{\Omega}^2(x)\tanh^2(x-X) dx.  
\end{align*}

\begin{figure}
	\includegraphics[scale=0.24]{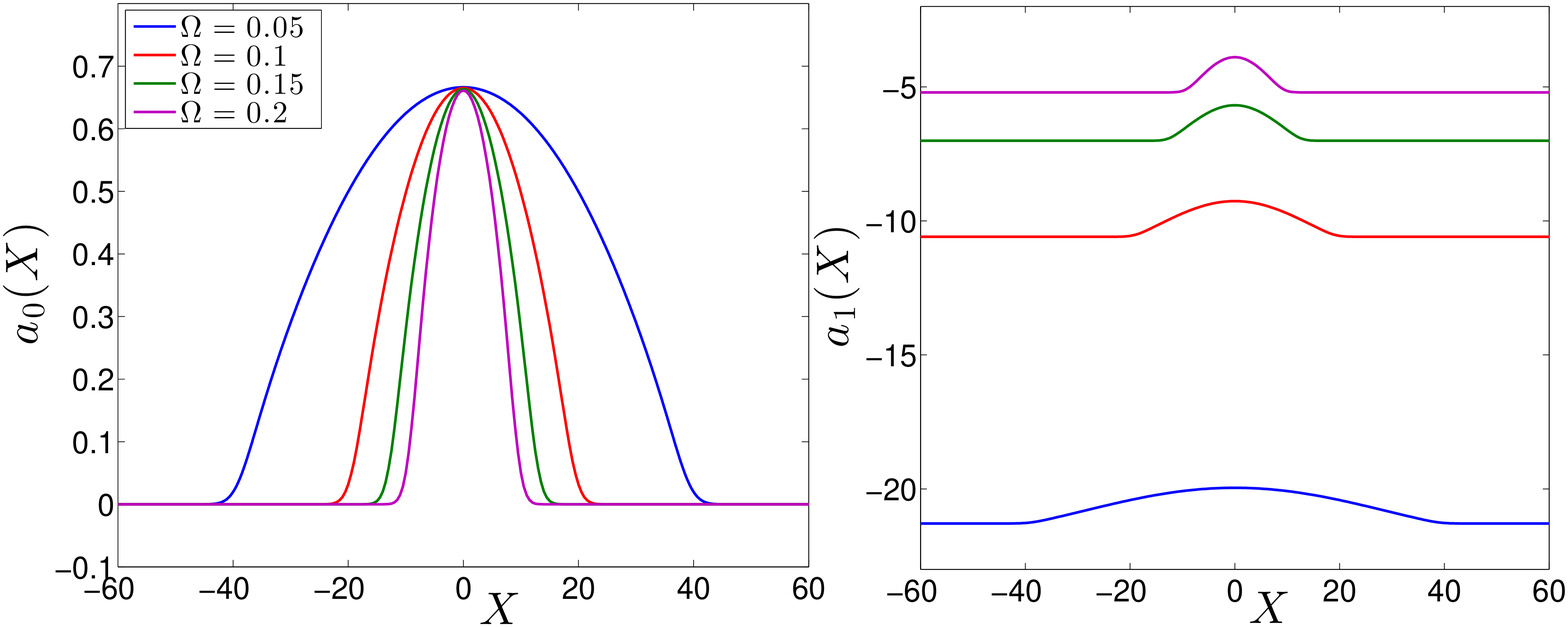}
	\caption{The left panel shows $a_0(X)$ and the right panel shows $a_1(X)$ as $\Omega$ varies: 0.05 (blue), 0.1 (red), $0.15$ (green), $0.2$ (purple). The inset plot on the right is the magnified view for $\Omega=0.15$.}
	\label{a0_a1_all}
\end{figure}

Since we have an extremely accurate numerical solution for $u_\Omega$, we compute the coefficients $a_0$ and $a_1$ and solve the Euler-Lagrange equation numerically. Fig. \ref{a0_a1_all} shows the plots of $a_0$ and $a_1$ as a function of $X$ for various values of $\Omega$. 
By applying the Euler-Lagrange equation, 
we obtain the dynamical evolution:
\begin{equation}\label{eq:EL}
\begin{aligned}
&\dot X = Y \\
&\dot Y = -\frac{1}{2}\frac{a'_0(X)}{a_0(X)}Y^2 - \frac{1}{2}\frac{a_1'(X)}{a_0(X)}.
\end{aligned}
\end{equation}
We solve these equations numerically by using the initial conditions $X(0)=x_0$ and $X'(0)=v_{\mathrm{in}}$ where $x_0$ is the initial position of the kink
and $v_\mathrm{in}$ is the initial velocity of the kink. We use MATLAB's
built-in fourth-order Runge--Kutta variable-step size solver {\tt ode45}
with built-in error control. Note that here, overdots denote
time derivatives.

\subsubsection{Connection to Numerical Results}
\begin{figure}[tbp]
\begin{center}
      \subfigure[]{{\includegraphics[width=0.32\textwidth]{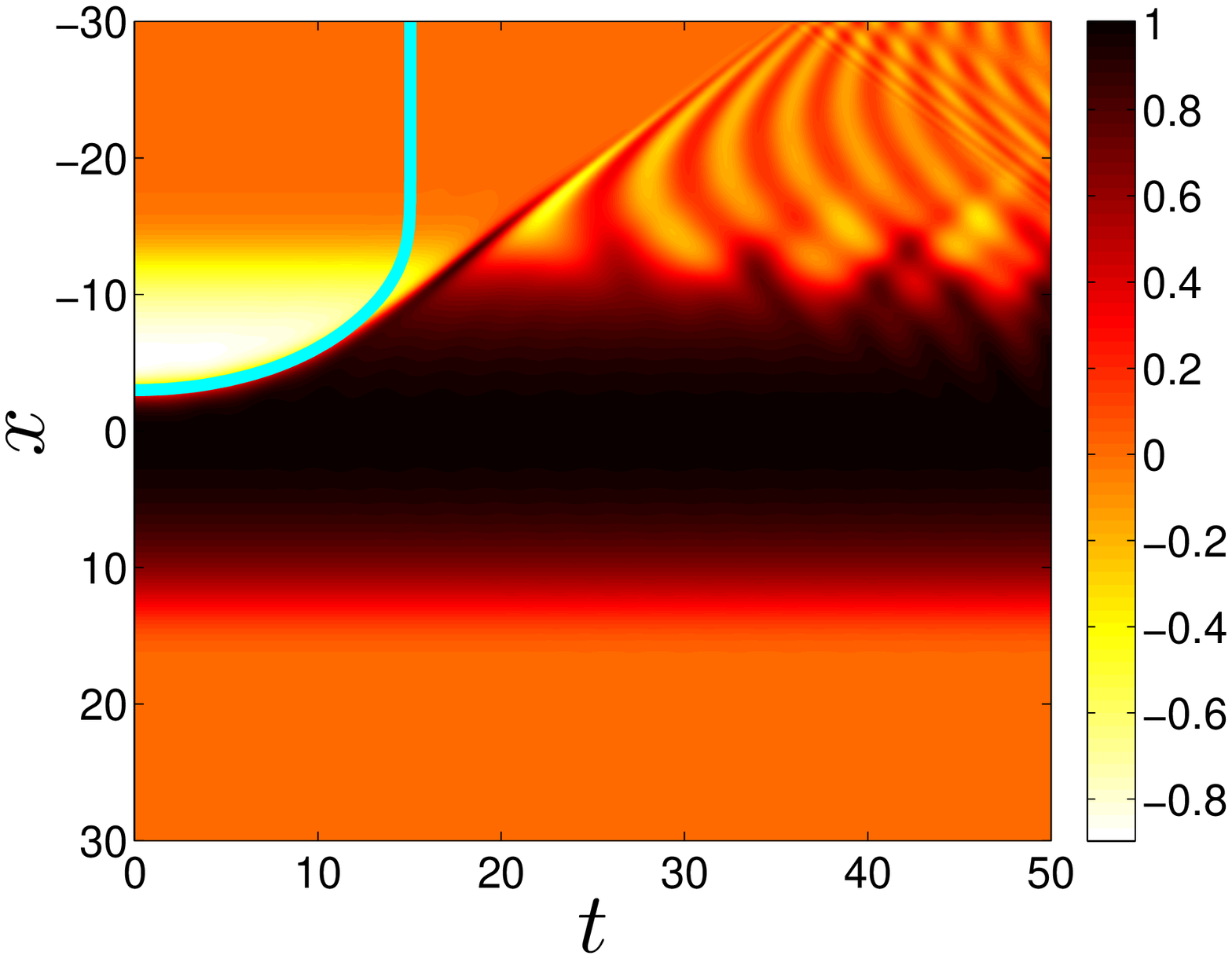}}}
      \subfigure[]{{\includegraphics[width=0.32\textwidth]{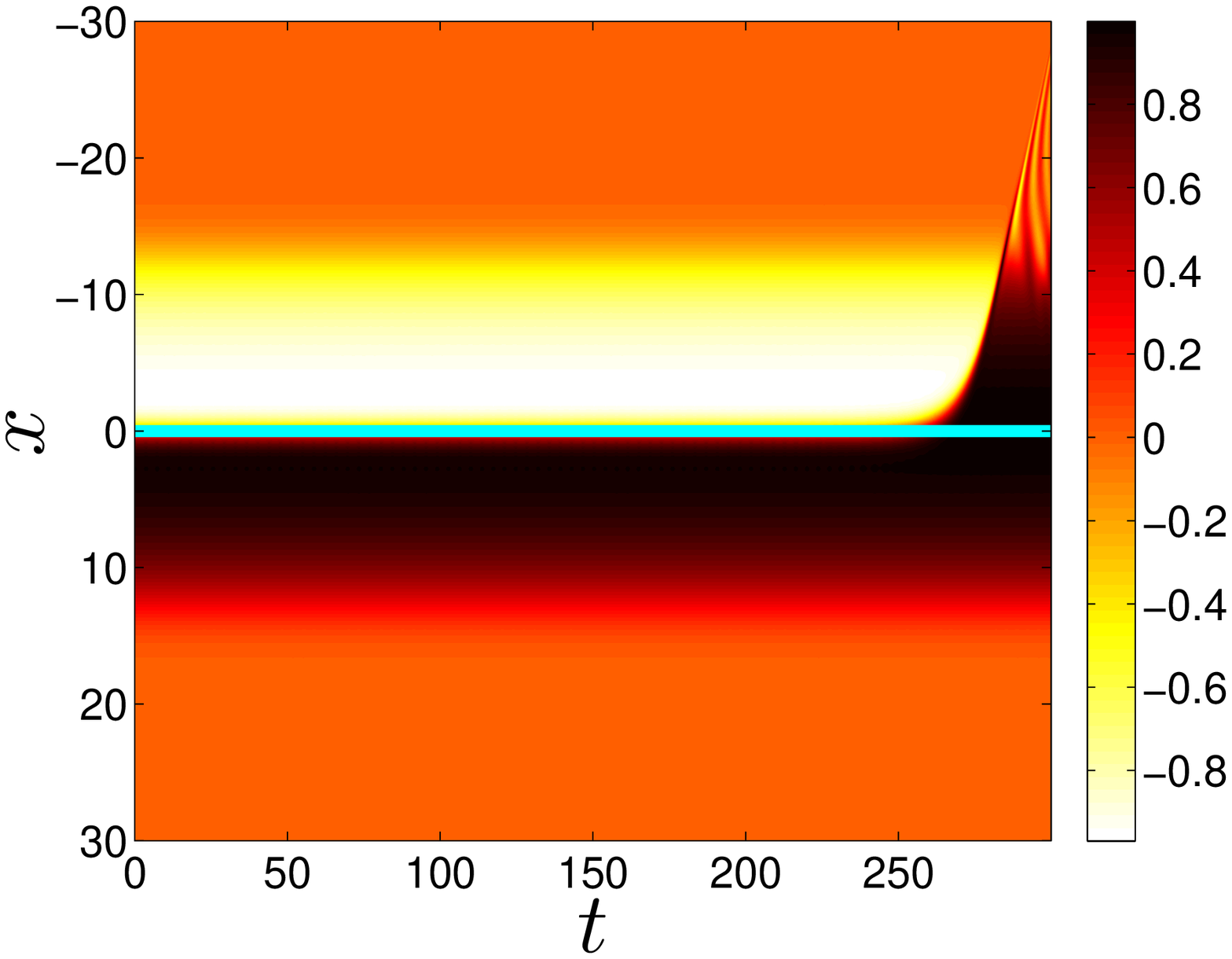}}}
       \subfigure[]{{\includegraphics[width=0.32\textwidth]{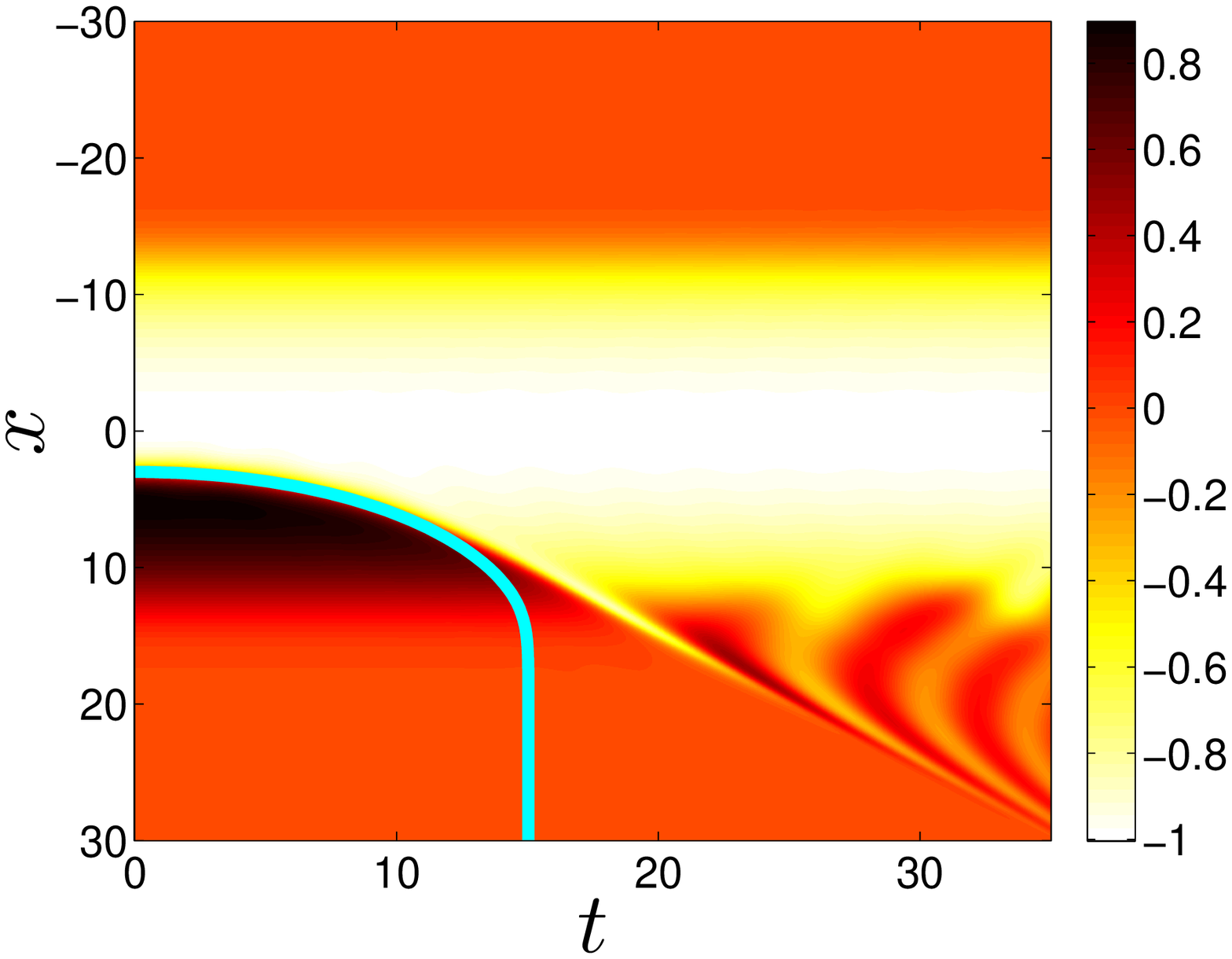}}}
  \end{center}
    \caption{ Overlays of the ODE solution $X(t)$ (in solid line) on top of PDE results for initially stationary ($v_\mathrm{in}=0$) kinks with (a) $x_0 = -3$ (b) $x_0=0$ (c) $x_0=3$. }
\label{ODE_PDE_single}
\end{figure}

We now compare our full numerical results (PDE) with the findings of the
CC ODE Method. 
The numerical CC computations show that when the initial velocity is zero ($v_\mathrm{in}=0$), with nonzero displacement ($x_0<0$ or $x_0>0$), the ODE results agree with PDE results quite well up to the time/location when/where the
kink is expelled from the system; from there on the kink cannot be
clearly discerned anyway, and hence it is not particularly meaningful
to seek to track it in the PDE or to match it to the ODE results.
Fig.~\ref{ODE_PDE_single}(a) and (c) show that when $x_0=-3$ and $x_0=3$, with zero initial velocity, both results mostly agree until $t \approx 11$
when the kink is expelled from the TF background.

In the case of  zero displacement ($x_0=0$) and zero velocity
($v_\mathrm{in}=0$) of Fig.~\ref{ODE_PDE_single}(b), both the ODE and the PDE show the kink
as residing on the saddle point for a long time interval. In the
ODE case, the absence of additional degrees of freedom and the conservation
of the energy does not allow the kink to depart from the unstable
equilibrium over the time horizons considered. In the infinite dimensional
system, the instability is manifested at $t \approx 250$.

\begin{figure}[tbp]
\begin{center}
      \subfigure[]{{\includegraphics[width=0.49\textwidth]{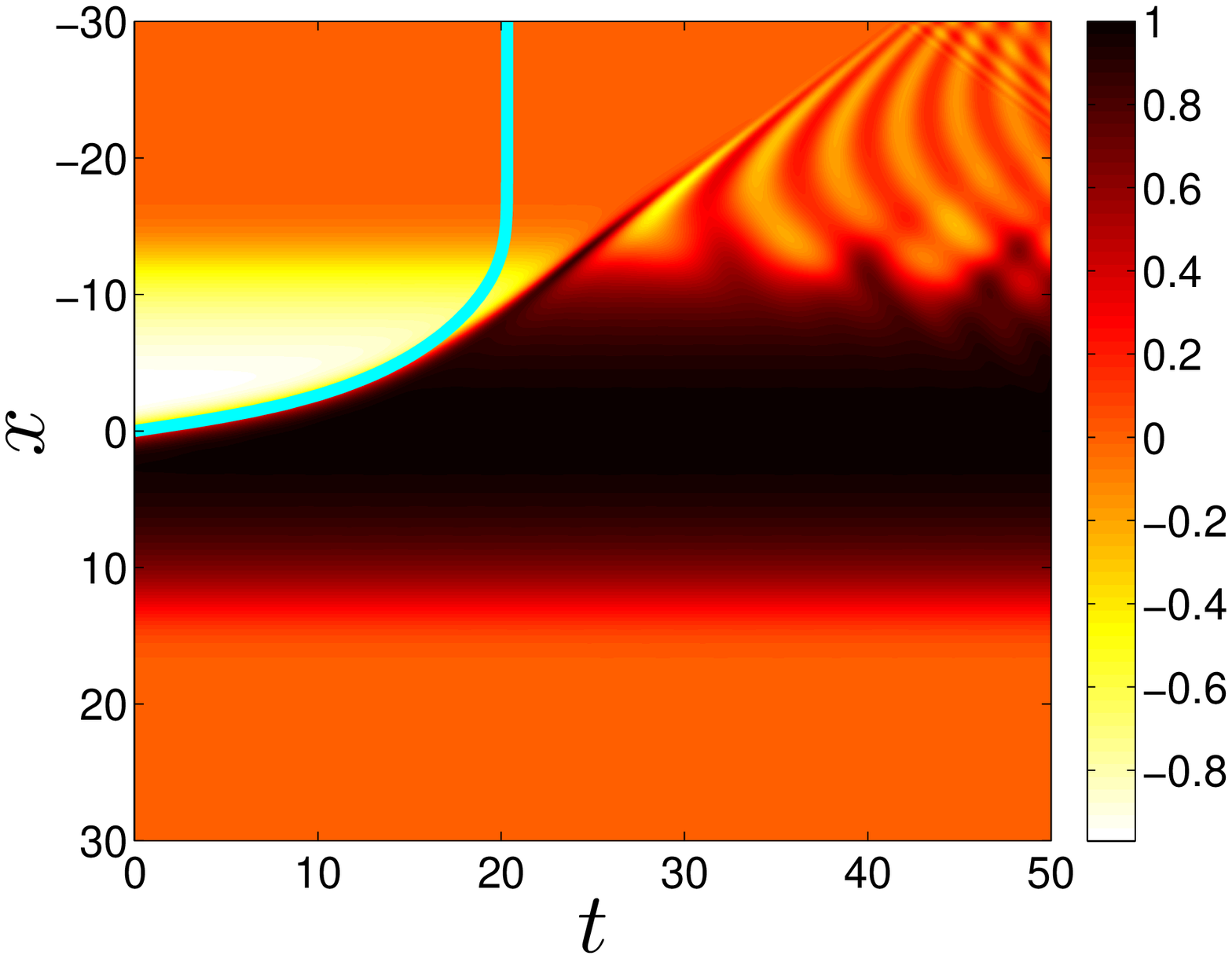}}}
      \subfigure[]{{\includegraphics[width=0.49\textwidth]{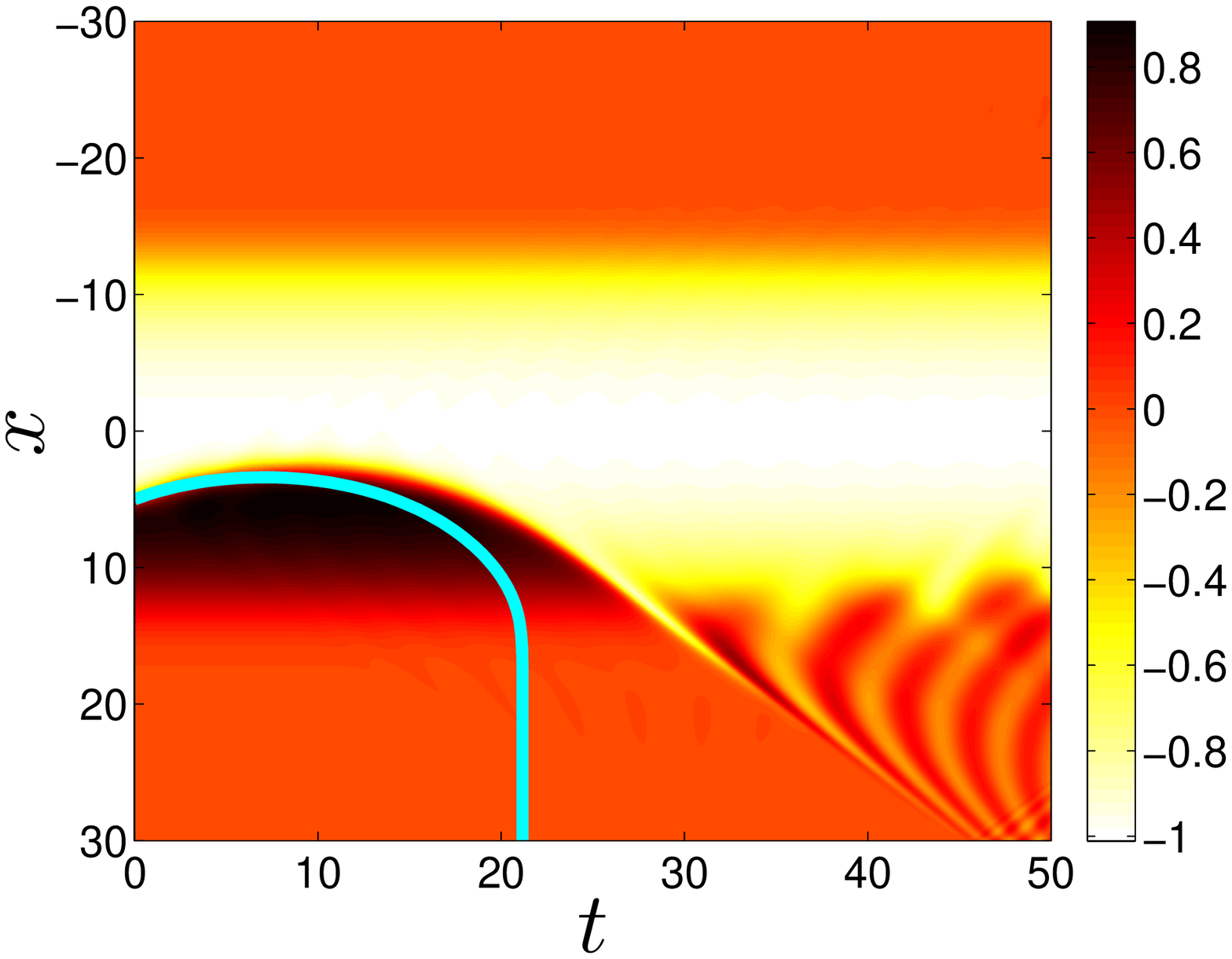}}}
  \end{center}
    \caption{ (a) A moving kink with $x_0=0$, $v_\mathrm{in}=0.2$, is
    shown again in comparison between ODE and PDE. (b) $x_0=5$, $v_\mathrm{in}=0.5$, A kink moving towards the origin with almost enough
    energy to break through the barrier at $0$. Both the ODE and PDE
    show it being reflected from the barrier at $x=0$. }
\label{ODE_PDE_single2}
\end{figure}

When the initial velocity is nonzero, we find similar results. The
ODE results agree with the PDE results until the ODE kink is expelled from the system.  Fig.~\ref{ODE_PDE_single2} (a) exhibits the ODE and PDE agreement until $t\approx 15$ when $x_0=0$ and $v_\mathrm{in}=0.2$. Fig.~\ref{ODE_PDE_single2} (b) exhibits this type of agreement until $t\approx 23$ when $x_0=-5$ and $v_\mathrm{in}=0.5$. In summary, we can infer that while the kink remains within the
region where the TF background is accurate, the CC method appears to yield an excellent qualitative
and a very good quantitative characterization of the resulting motion.

 \section{Kink-Antikink Solutions}

Armed with our understanding of the single kink energy landscape, we now turn to case of a kink-antikink pair. Such a state can in
fact be {\it stationary} in our model, although it is never genuinely
stationary in the case of $\Omega=0$. The reason is fairly similar
to the corresponding existence of a stationary state in a nonlinear
Schr{\"o}dinger (NLS) model with a trap
analyzed in detail, e.g., in~\cite{coles}.
However, there is a {\it crucial} difference. In the latter
case, the trap has a restoring contribution, while the
dark solitons (the kinks) of the NLS repel each other. Here, the situation is
reversed: namely, as we saw in the previous section,
the trap tends to expel the kink, while (as is known from the
study of kink-antikink interactions in the homogeneous $\Omega=0$
case) the kink and antikink attract each other ~\cite{anninos,campbell},
i.e., the role of both
contributions is reversed. As a result, the equilibrium point
instead of being a center as it is in the case of NLS, here, again,
it is a saddle. Indeed, if the kink pair is separated by a small
distance, then the interaction prevails and forces the kinks to
approach each other. On the other hand, a large separation
leads to a dominant effect due to the trap, and the kink-antikink
pair are pushed apart.

\begin{figure}[h]
	\centering 
		\includegraphics[width=\textwidth]{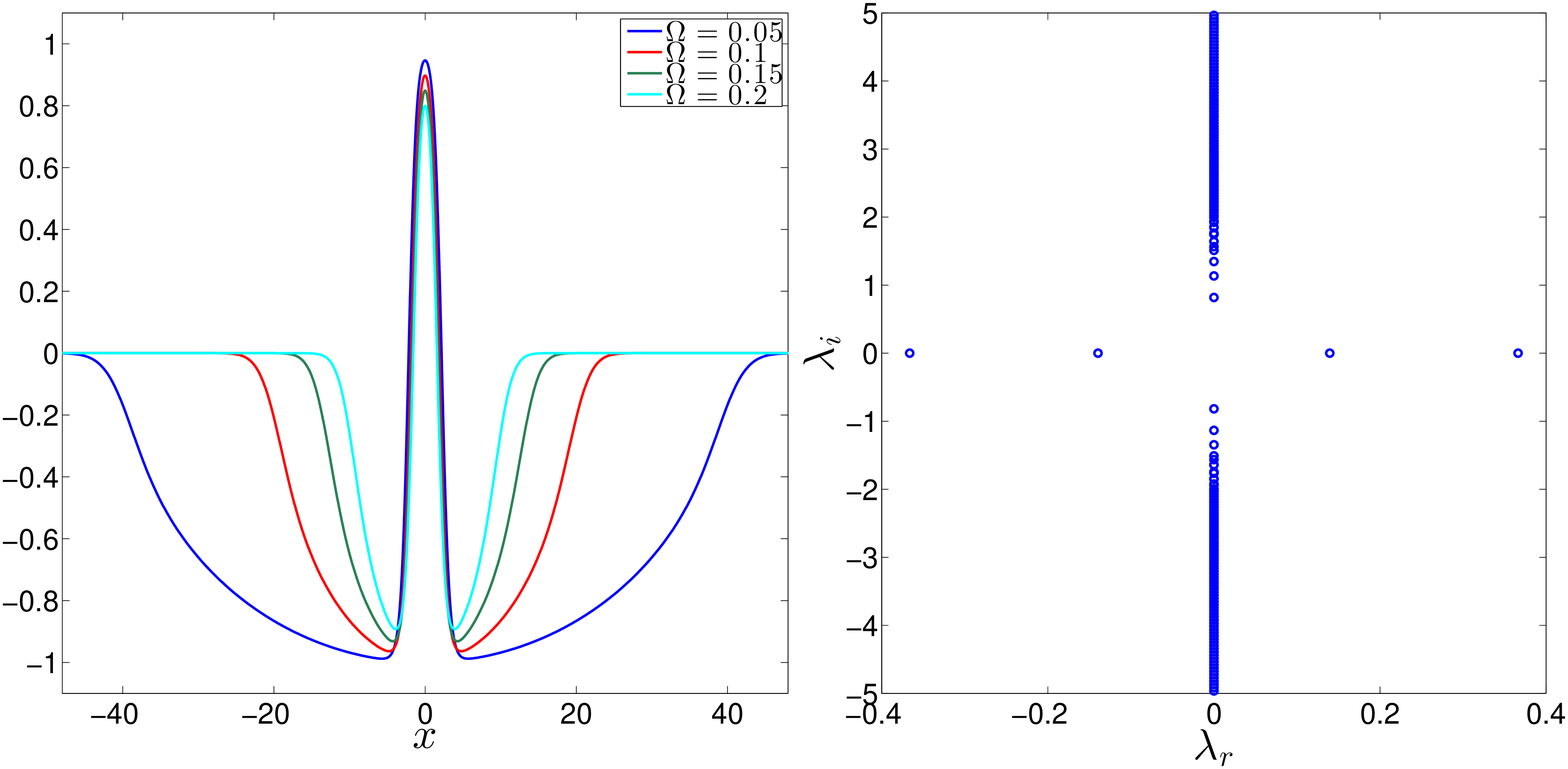}

	\caption{ Kink-antikink stationary solutions for
        different values of $\Omega$. For $\Omega=0.05$, the equilibrium
        sepration is $x_0=2.173$, for $\Omega=0.1$, it is $x_0=1.8669$, for
        $\Omega=0.15$, it is $x_0=1.6907$, while for $\Omega=0.2$,
        it is $x_0=1.5661$. The spectral plane reveals two real eigenvalue
        pairs, again signaling the instability when $\Omega=0.15$ (the
        case shown), as well as for all other values of $\Omega \neq 0$.} 
	\label{kink-antikink}
\end{figure}

The static kink-antikink solution of (\ref{fullpde})
can be numerically approximated by 
\begin{equation}u_{K,\tilde{K}} (x) \approx  u_{\Omega}(x) (\tanh (x+x_0)-\tanh (x-x_0)-1).
\label{static_kink-antikink}
\end{equation}
where $x_0$ is unique and dependent on $\Omega$. We define that particular $x_0$ as $x_\mathrm{cr}$, since it operates as a critical value separating
between expulsion from and attraction of the kink and antikink to the center. 
For various values of $\Omega$, the numerically calculated
stationary kink-antikink solutions are shown in Figure \ref{kink-antikink}.
As usual, the right panel of the figure identifies the spectrum of the linearized operator about the static (in this case kink-antikink) solution.
Here, we get two pairs of real eigenvalues, indicating instability.
The pattern of the spectral problem is thus transparent: for one kink,
there is a single unstable mode, for two kinks, two such modes and
so on. Another way to state this is that  all the negative energy eigenstates
of~\cite{coles} pertaining to the normal modes of vibration of the kinks
in the real-field version of the model lead to unstable eigendirections.

\subsection{Numerical Results (PDE)}
We can obtain moving kink-antikink solutions by picking the initial conditions 
\begin{align*}u(x,0)&=u_{\Omega}(x) (\tanh(\gamma (x+x_0))- \tanh(\gamma (x-x_0))-1); \\
u_t(x,0)&=-v \gamma u_{\Omega}(x) (\sech^2(\gamma(x+x_0))+ \sech^2(\gamma(x-x_0)));
\end{align*}
where $\gamma=1/\sqrt{1-v^2}$, $\pm x_0$ are 
the initial positions and $\pm v$ are the velocities of the moving kink
and antikink.
Once again,
we utilize the (no longer representing an invariance) Lorentz boost
and find it to be an efficient method of producing moving kinks-antikinks.

\begin{figure}[tbp]
\begin{center}
 \subfigure[]{{\includegraphics[width=0.49\textwidth]{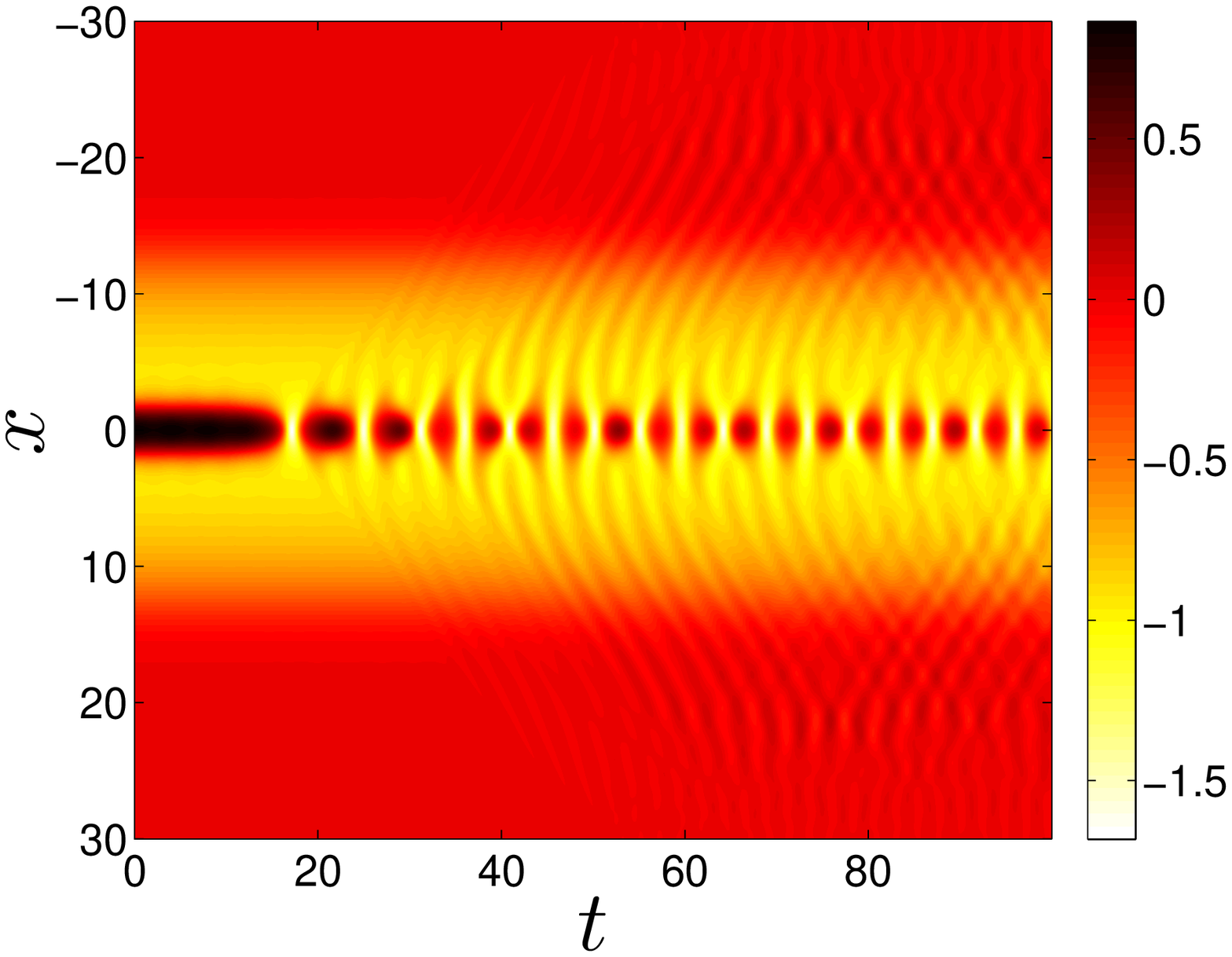}}}
      \subfigure[]{{\includegraphics[width=0.49\textwidth]{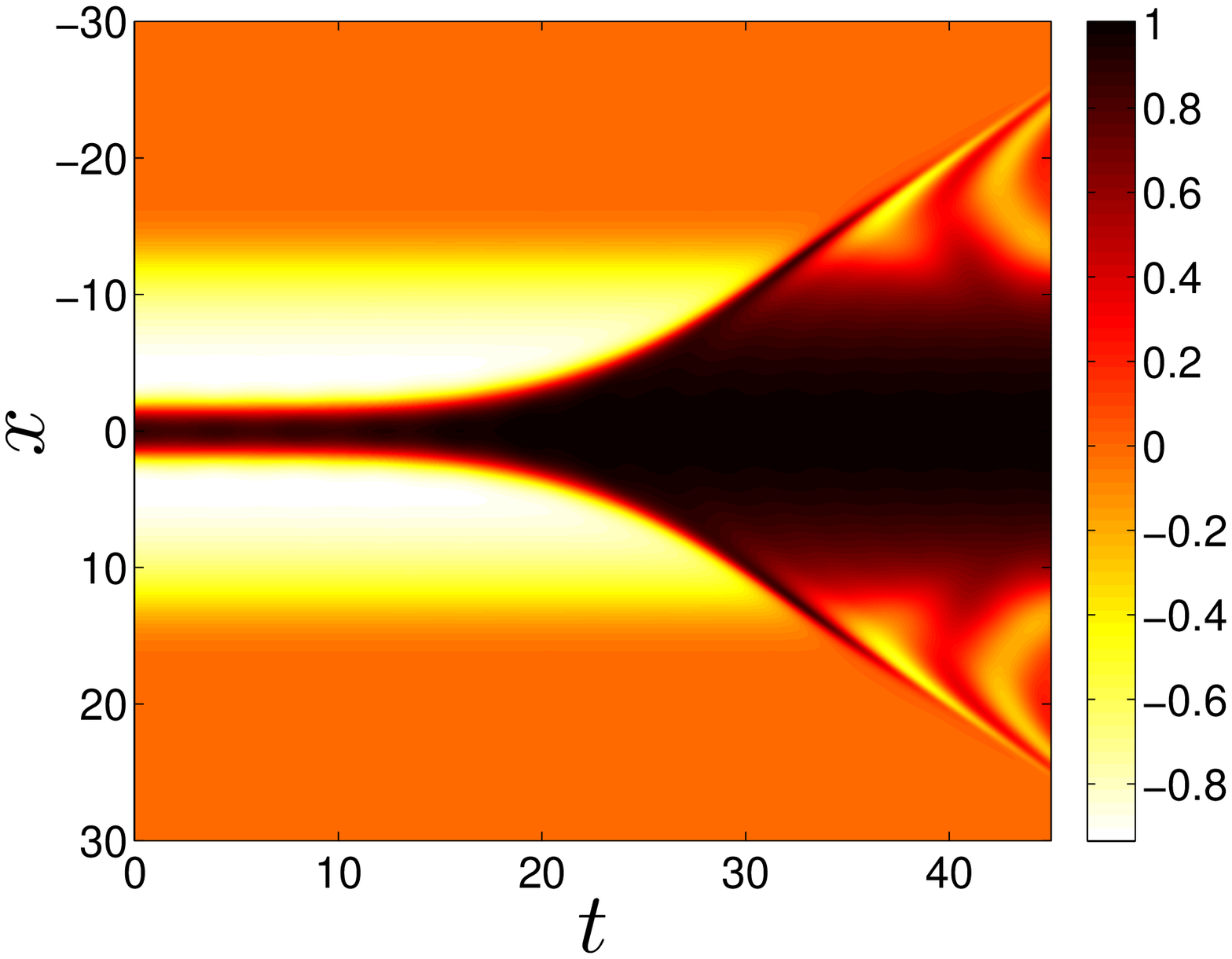}}}
  \end{center}
   \caption{(a) ($x_0=1.69$) a kink-antikink system with zero initial velocity for $x_0 < x_\mathrm{cr}$ forms a bound state. (b)($x_0=1.7$), a zero-velocity configuration with $x_0>x_\mathrm{cr}$ exhibits expulsion.}
    \label{before_after_xc}
\end{figure}

Fig.~\ref{before_after_xc} shows that when the velocity of the kink and antikink is initially zero, then for any $x_0<x_\mathrm{cr}\approx 1.6907$ (for $\Omega=0.15$), the kink-antikink system forms a bound state. For any $x_0>x_\mathrm{cr}$, the system exhibits expulsion. This confirms our spectral understanding
of the relevant equilibrium representing a saddle point. Now the unstable
manifold consists of two eigendirections, namely the in-phase and out-of-phase
motion of the pair, although here only the out-of-phase mode is excited.

\begin{figure}[tbp]
\begin{center}
      \subfigure[]{{\includegraphics[width=0.49\textwidth]{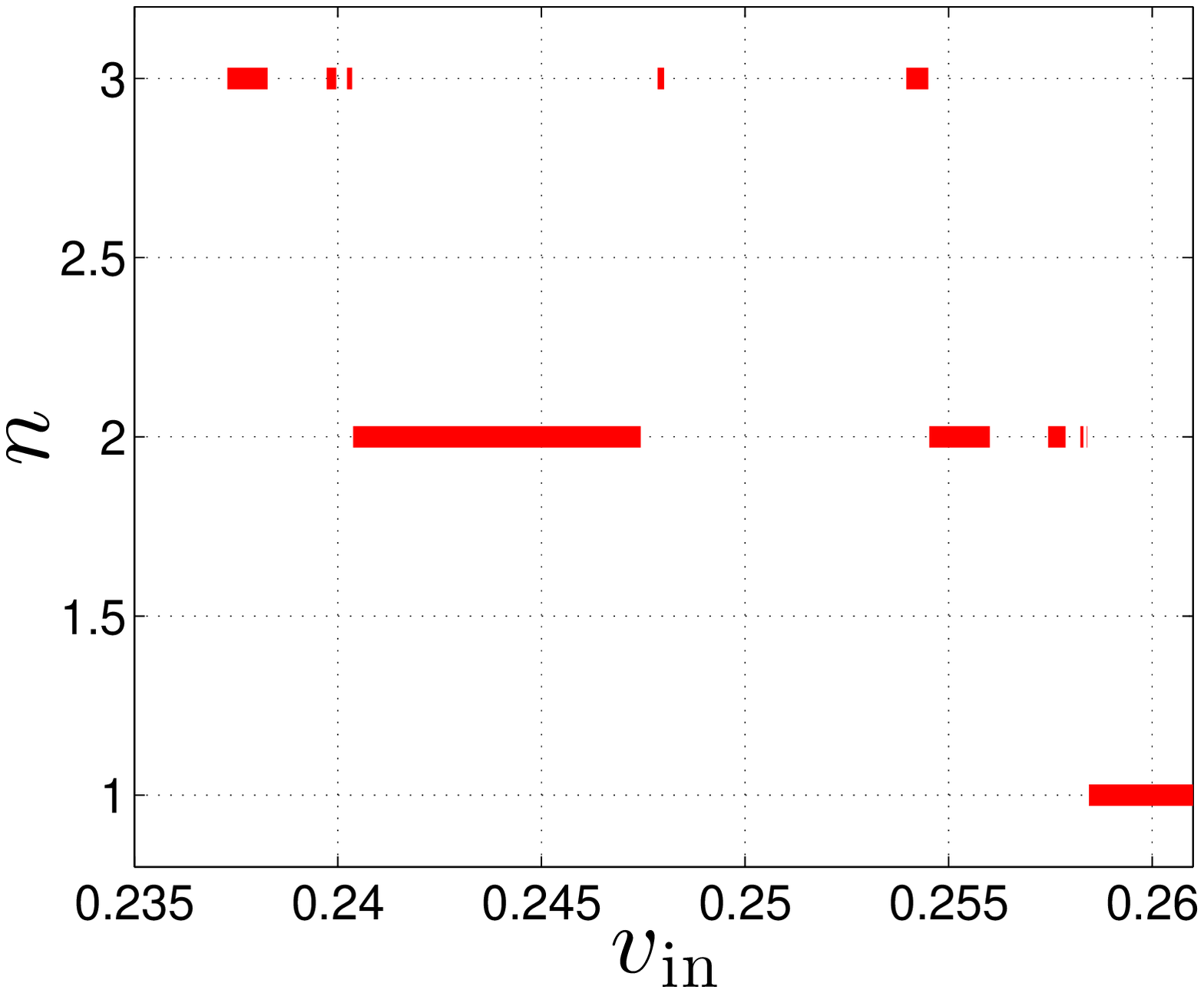}}}
      \subfigure[]{{\includegraphics[width=0.49\textwidth]{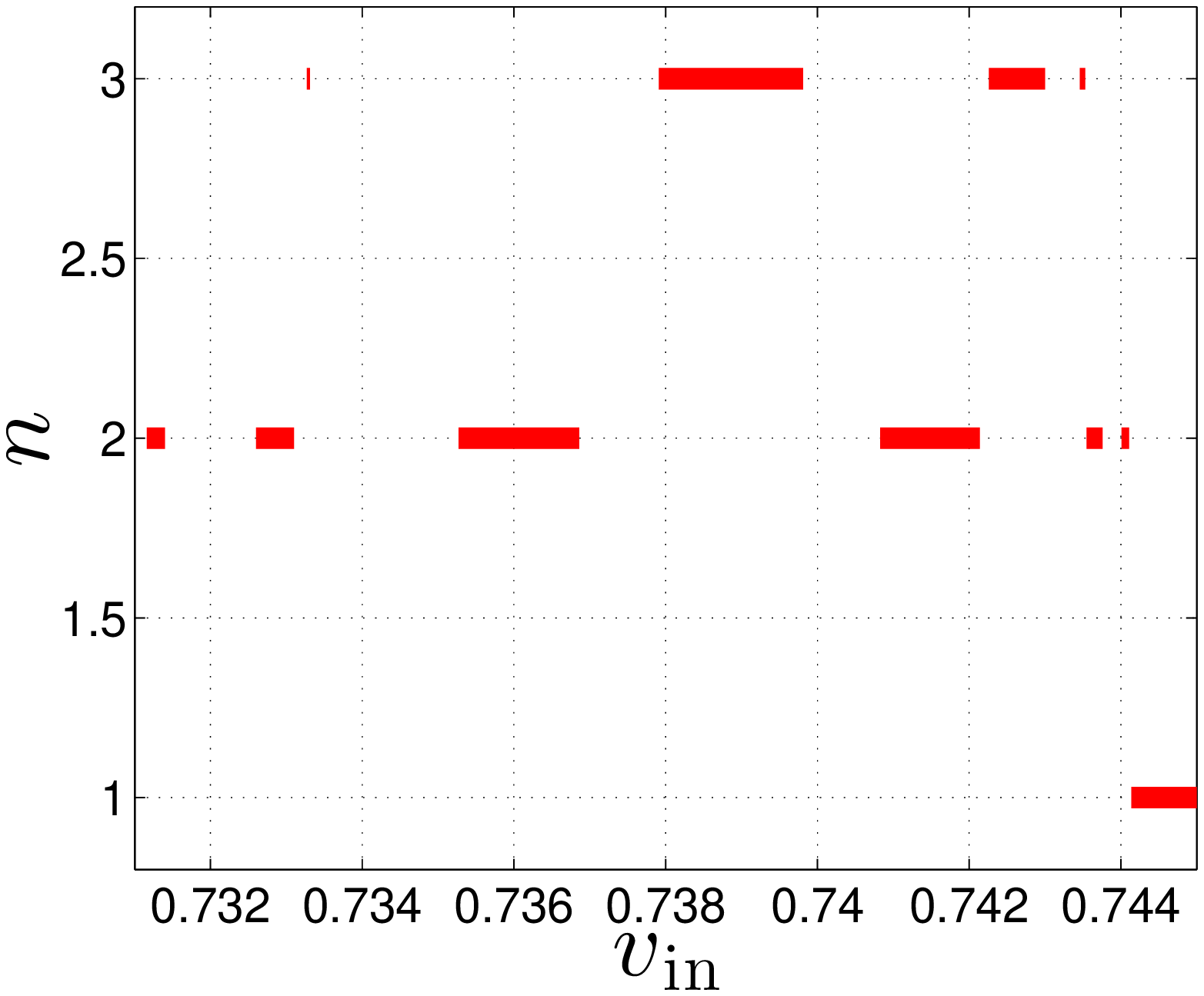}}}
  \end{center}
   \caption{ Plots of number of bounces vs. initial velocity with (a) $x_0=1.4$ for $v_\mathrm{in} \in (0.235, 0.261)$ and (b) $x_0=7$ for $v_\mathrm{in} \in (0.731,0.745) $. As in the homogeneous case, we observe ranges for which multiple bounces occur.}
    \label{windows_Omega_015}
\end{figure}

When $\Omega=0$, the kink-antikink system is relatively insensitive to changes in the initial position $x_0$ (provided that the kinks are not too close).
This, however, changes in the presence of the
parabolic trapping term. For $x_0$ sufficiently small, the kink and
antikink approach each other and collapse into a bound state for small input velocities, even with zero initial velocity. For $x_0<x_\mathrm{cr}$ and initial velocities sufficiently large we observe the typical behaviors that
are familiar from the homogeneous case~\cite{campbell,anninos}: reflection,
$n$-bounce windows (i.e., windows where the kinks escape each
other's attraction upon $n$ collisions), and  {``bion/breather" \cite{eilbeck,campbell} formation (i.e.,
emergence of a bound state that oscillates in time and gradually decays). In fact, the types of resonance windows observed in the homogeneous case~\cite{campbell,anninos,goodman2}}
persist in the inhomogeneous case and present a similar
type of  pattern. A visualization of this can be seen in Fig.~\ref{windows_Omega_015}. In this figure and in subsequent figures, we opt not to plot velocities for which
four or more bounces occur, as the windows become exceedingly narrow
for this many bounces and it is not practical to display them
together with one- and two-bounce windows. Just as when $\Omega=0$, the regions in between resonance windows correspond to ``bion" formation (numerically
these are treated as events with many bounces, so they are omitted from the
figures).

For $x_0 > x_\mathrm{cr}$,
the presence of the saddle point of the energy once again imposes
a barrier that the kinks need to overcome in order to interact.
Thus, in this case, there is a critical velocity associated to $x_{cr}$, which we denote by $v_\mathrm{cr}$, such that whenever $x>x_\mathrm{cr}$ and $v < v_\mathrm{cr}$ the kinks do not have enough kinetic energy to reach the origin, so they stop (at a turning point) before they reach it. Then they recede
towards spatial infinity without ever having collided.

If both $x>x_\mathrm{cr}$ and $v > v_\mathrm{cr}$, then the kinks will have
enough kinetic energy to overcome the barrier, reach the origin and collide,
and then we observe the typical phenomena associated with the
$\phi^4$ model, including most notably the existence of multi-bounce windows.
These can be visualized in Figs. \ref{windows_Omega_015} and \ref{fig:windows_Large_Omega} with both small and large separation $x_0$, respectively. The velocity interval data that was used to create these figures is contained in Tables 1--4 in Appendix A.

\subsection{Collective Coordinate Approach (ODE)}

 To complement our PDE simulations, we formulate a collective coordinate model for a kink-antikink configuration. As before, our aim is to reduce the full PDE with infinitely many degrees of freedom to a simple model with two degrees of freedom in a way similar to what has been previously done
 in the homogeneous $\phi^4$ case~\cite{campbell,anninos,goodman2,goodman}. 
To account for the internal modes of the kink and antikink, we introduce a new unknown $A(t)$, representing the amplitude of the internal mode. We use only one function for both modes since the collision picture is symmetric between kink and antikink, so we expect the amplitudes of the kink mode and antikink mode to
coincide if the initial configuration is symmetric. 
Our corresponding ansatz in this case will be:
\begin{equation}\label{ansatz}
u(x,t)=u_{\Omega}(x)(u_0(x+X)-u_{0}(x-X)-1)+A(t)(\chi_1(x+X)-\chi_1(x-X)) 
\end{equation}
where $u_{0}(x)=\tanh(x)$, and $\chi_1$ is an eigenfunction corresponding to the smallest positive eigenfrequency that satisfies the linearization equation for a single kink.

\begin{figure}[h]
\begin{center}
	\includegraphics[width=0.5\textwidth]{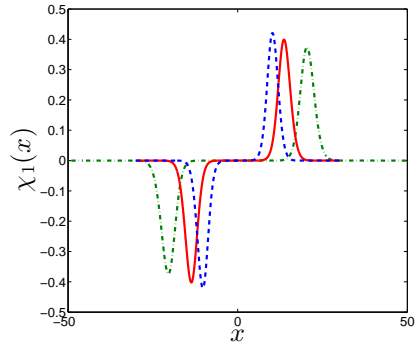}
	
 \caption{Plot of the eigenfuction used in the collective coordinates method. Green dashed-dots correspond to $\Omega=0.1$, red solid corresponds to $\Omega=0.15$ and the dashed blue corresponds to $\Omega=0.2$.}
	\label{eigenvectors}
\end{center}
\end{figure}

The particular eigenfunction we chose is given in Figure \ref{eigenvectors} for three different $\Omega$ values.

 Surprisingly, these modes are related
 chiefly to the background state $u_{\Omega}$ rather than to the internal vibrational mode of the kink.
 More specifically, they are localized near $x_s$.
 In the homogenous model, the only nonzero discrete eigenfrequency corresponds to the internal vibrational mode, and it is this mode that is used for the collective coordinate ansatz. For our model, the mode with the smallest eigenfrequency results in a better match of collective coordinate and PDE results than when the (analogue of the homogeneous case kink) internal
  mode is used. This may be due to the former mode incorporating more
  adequately the role of the potential-induced background, although,
  admittedly, this issue warrants further theoretical investigation.

\begin{figure}[h]
\begin{center}
	\includegraphics[width=0.32\textwidth]{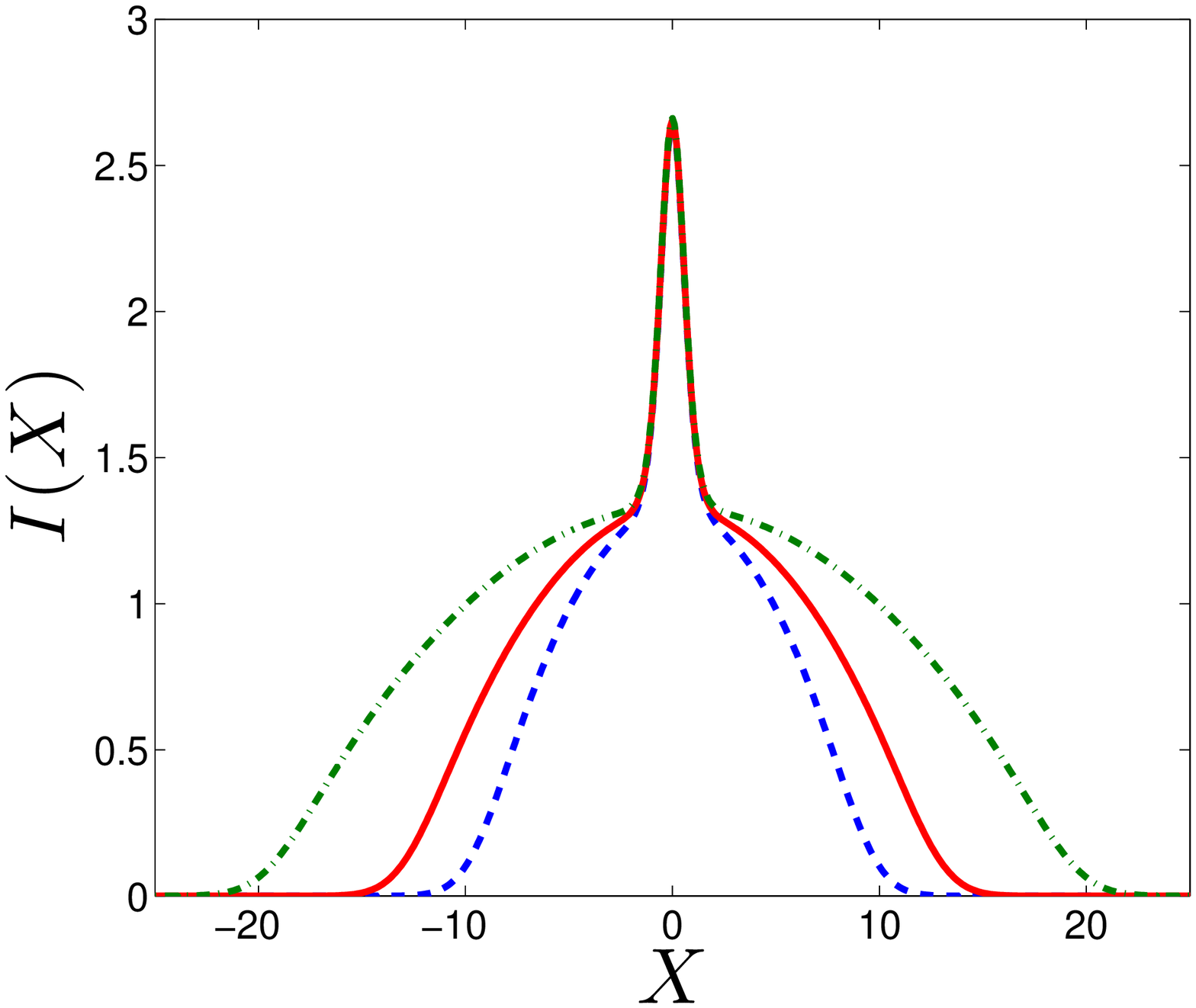}
	\includegraphics[width=0.32\textwidth]{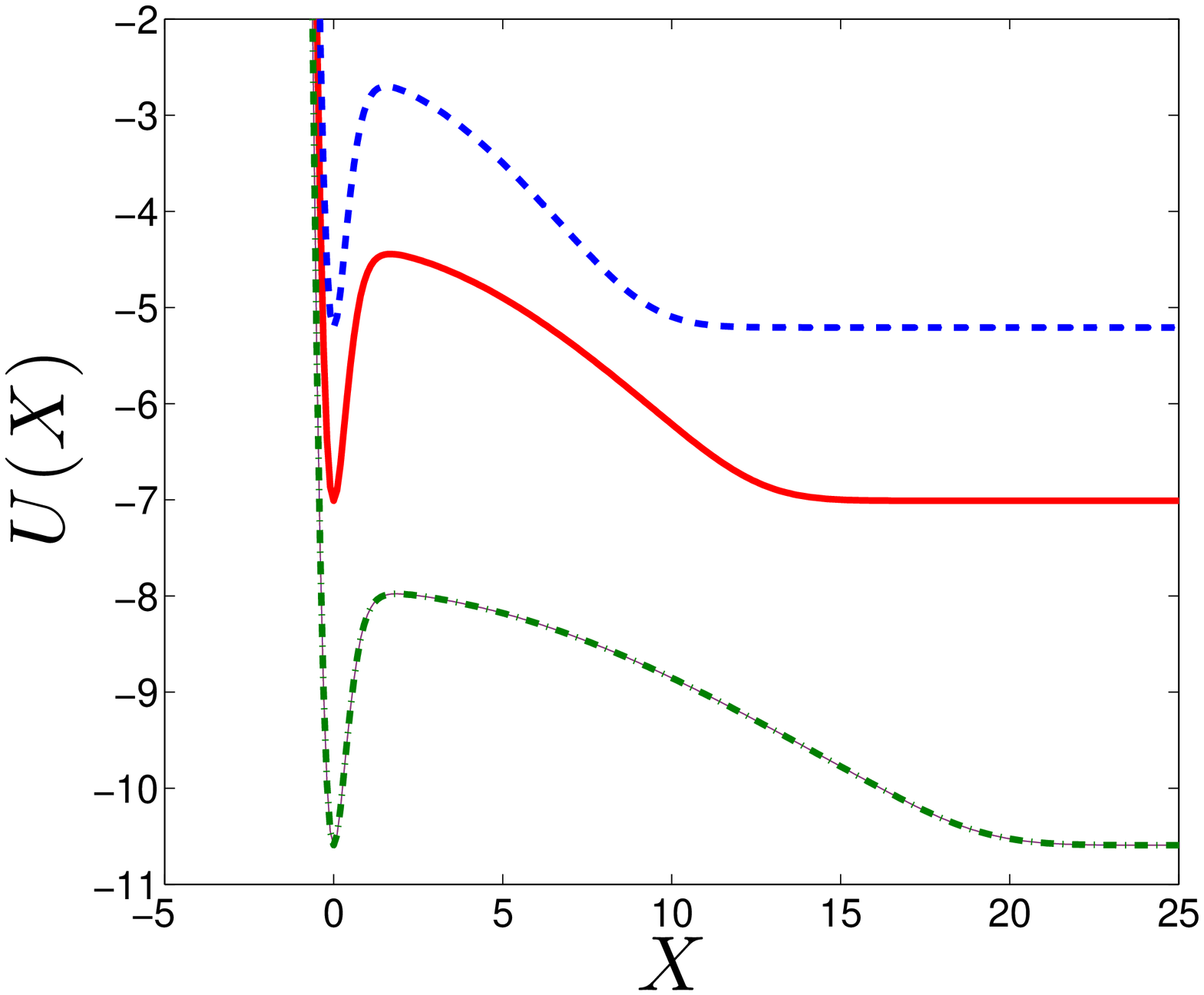}
	\includegraphics[width=0.32\textwidth]{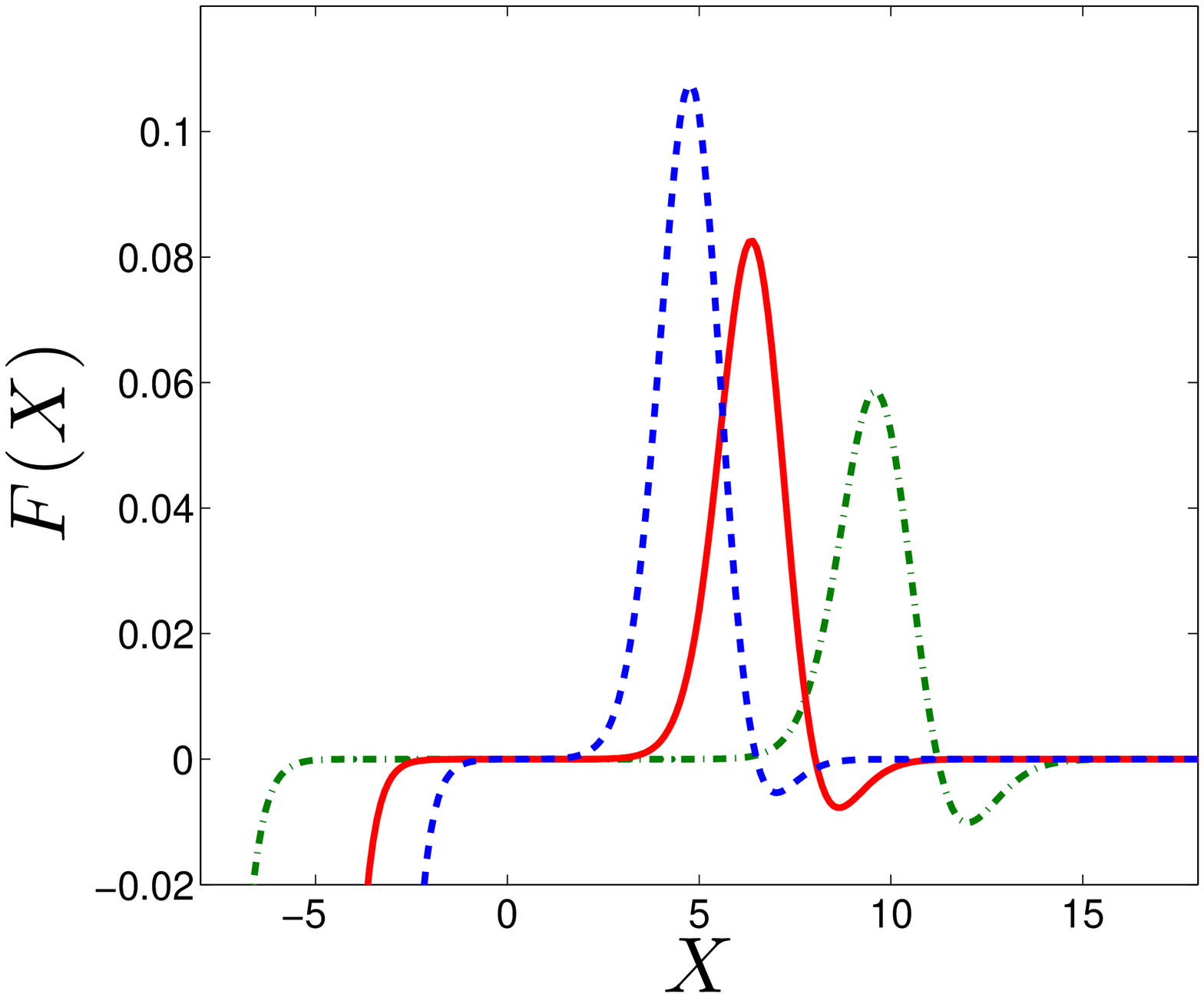}	
	\caption{Plots of the functions $I(X)$ (left), $U(X)$ (middle)
  and $F(X)$ (right) as a function of the position $X$ of the kink's
center, evaluated through the method of collective coordinates, as discussed in the text. Green dashed-dots correspond to $\Omega=0.1$, red solid corresponds to $\Omega=0.15$ and the dashed blue corresponds to $\Omega=0.2$.}
	\label{I_U_F}
\end{center}
\end{figure}

\begin{figure}[h]
\begin{center}
	\includegraphics[width=0.32\textwidth]{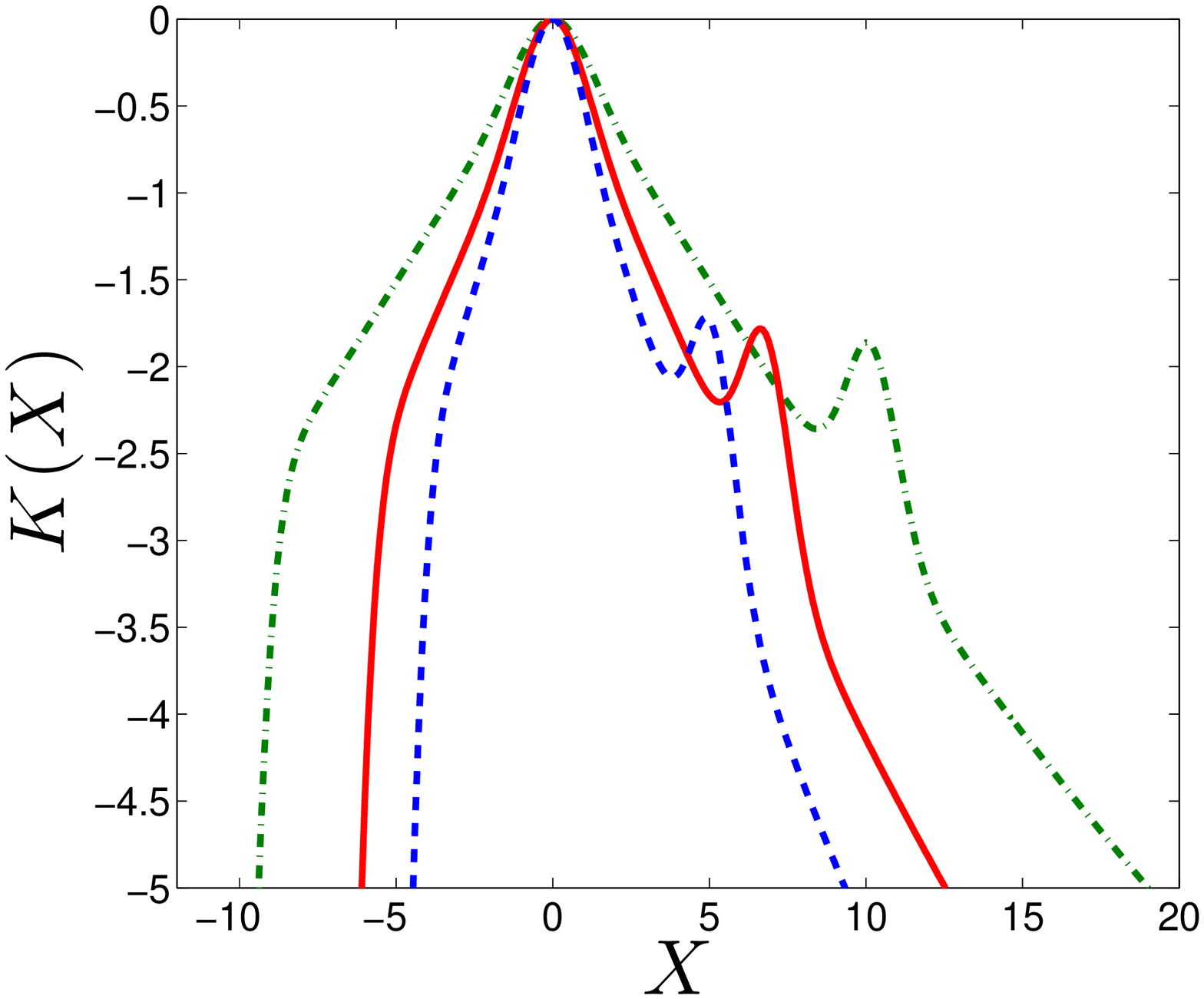}
	\includegraphics[width=0.32\textwidth]{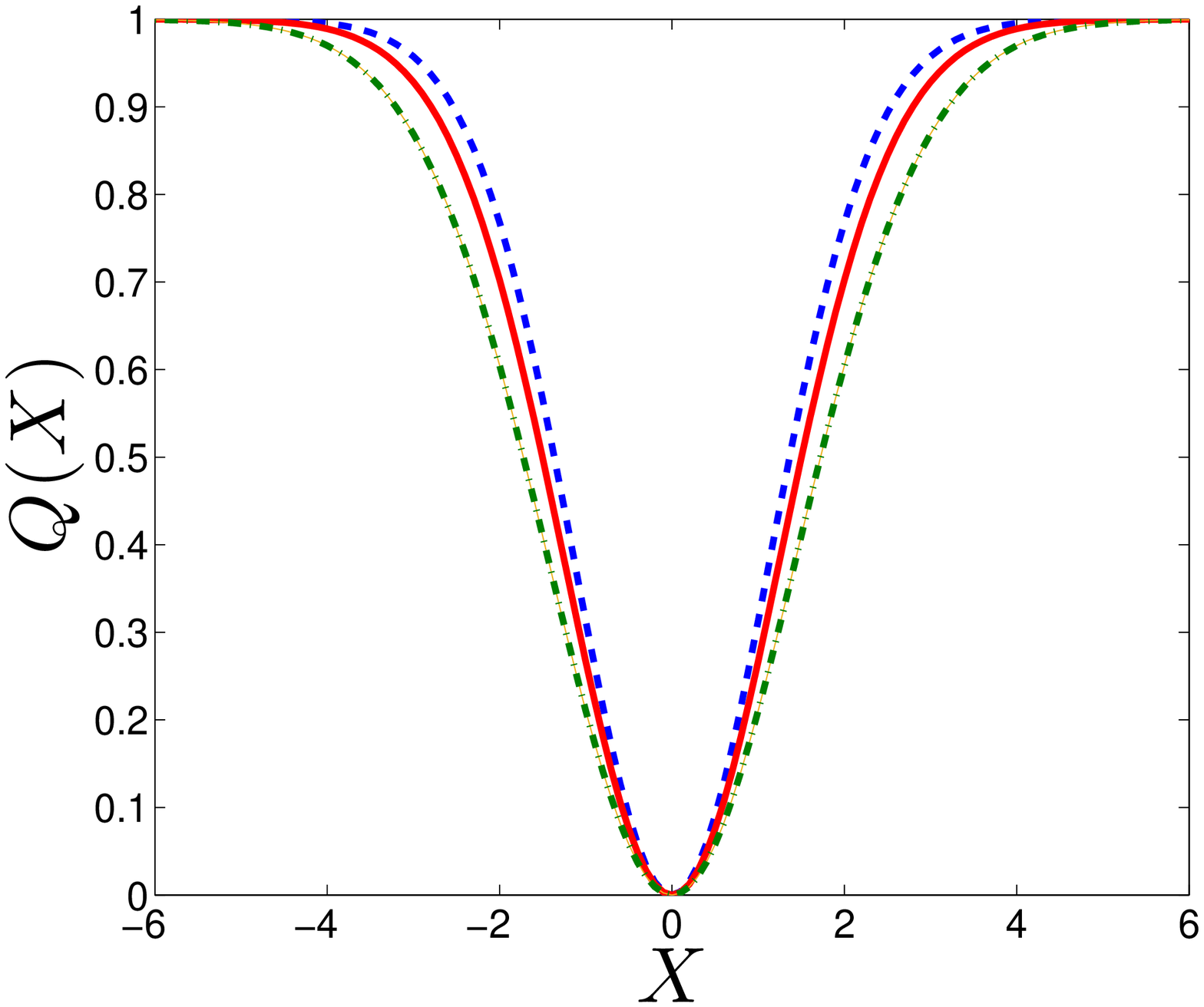}
	\includegraphics[width=0.32\textwidth]{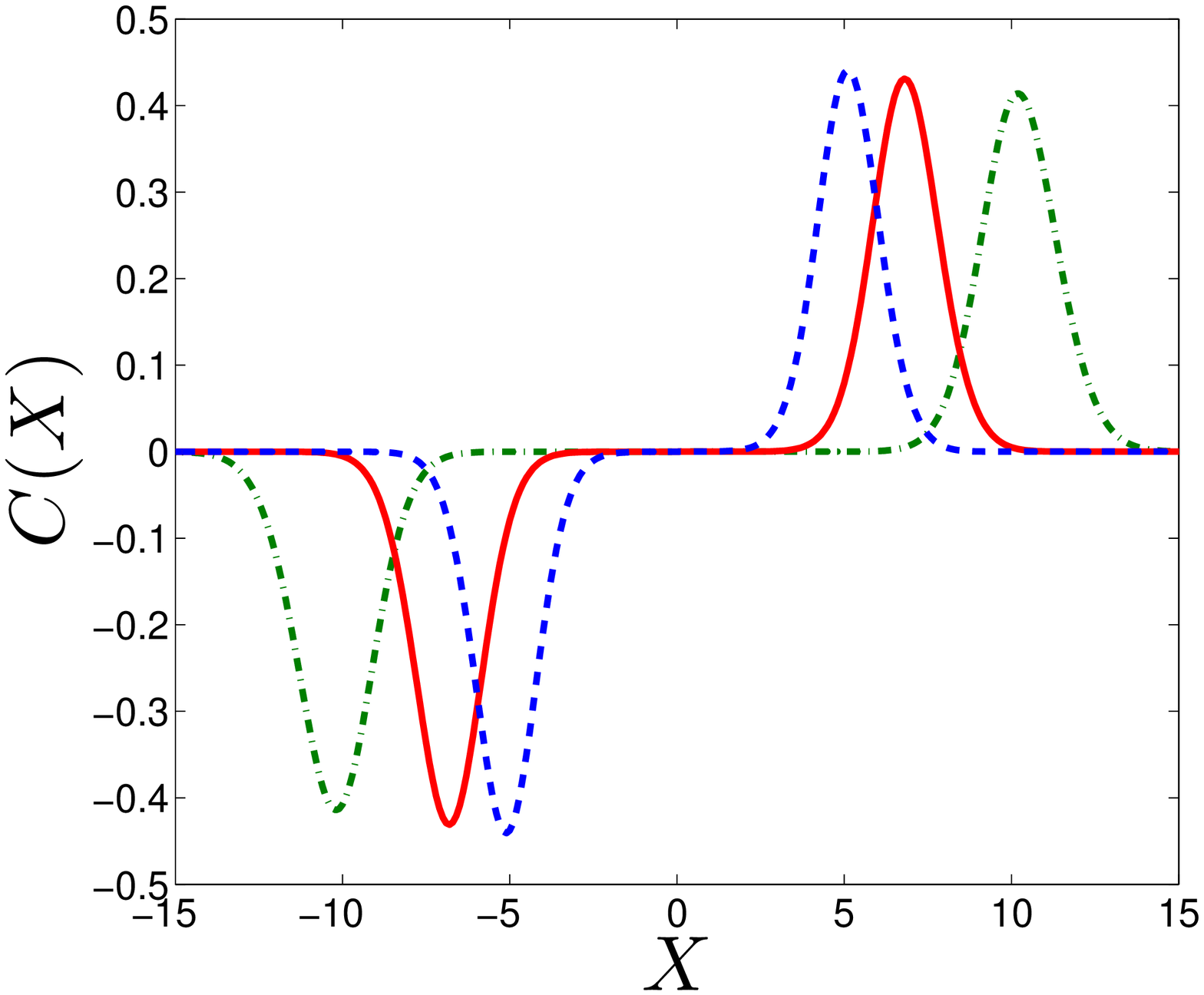}		
	\caption{Plots of the functions $K(X)$ (left), $Q(X)$ (middle)
  and $C(X)$ (right) as a function of the position $X$ of the kink's
center, evaluated through the method of collective coordinates, as discussed in the text. Green dashed-dots correspond to $\Omega=0.1$, red solid corresponds to $\Omega=0.15$ and the dashed blue corresponds to $\Omega=0.2$.}
	\label{K_Q_C}
	\end{center}
\end{figure}

As proposed in \cite{sugiyama}, we work on a reduced effective Lagrangian that captures the fundamental features. The derivations of the formulas are in Appendix B.
\begin{equation}\label{lag1}
L(X,\dot{X}, A, \dot{A})=I(X)\dot{X}^2-U(X)+2F(X)A + K(X)A^2 + Q(X)\dot{A}^2 + 2C(X)\dot{A}\dot{X}.
\end{equation}
Figs.~\ref{I_U_F} and~\ref{K_Q_C} show the collective coordinate coefficients as functions of $X$ for various values of $\Omega$.

The Euler-Lagrange equations in this case with respect to $X$ and $A$
give
\begin{equation}\label{EL_ode} 
\begin{aligned}
C(X)\ddot{X}+Q(X)\ddot{A} = -&Q'(X)\dot{X}\dot{A}-C'(X)\dot{X}^2+F(X)+K(X)A\\
2I(X)\ddot{X}+2C(X)\ddot{A} = -&I'(X)\dot{X}^2-U'(X)+2F'(X)A\\
&+K'(X)A^2+Q'(X)\dot{A}^2
\end{aligned}
\end{equation}

Here, overdots denote
time derivatives. We solve these equations numerically by using the initial conditions $A(0)=0$, $\dot{A}(0)=0$, $X(0)=x_0$ and $\dot{X}(0)=v_{\mathrm{in}}$ where $x_0$ is the initial half-distance between the kink and the antikink, and $v_\mathrm{in}$ is the initially
prescribed speed of the kink (and anti-kink).
{Note that this is the reduced system where the higher order
  terms of $A(t)$ and $\dot{X}(t)$ are ignored. We make this assumption  based on the previous work \cite{sugiyama}, that higher terms of $A$ and $\dot{X}$ can be dropped because they are small. The expressions involving integrals which are $X$-dependent in Eqs.~(\ref{EL_ode}) are computed by numerical integration.

\begin{figure}[tbp]
\begin{center}
      \subfigure[]{{\includegraphics[width=0.49\textwidth]{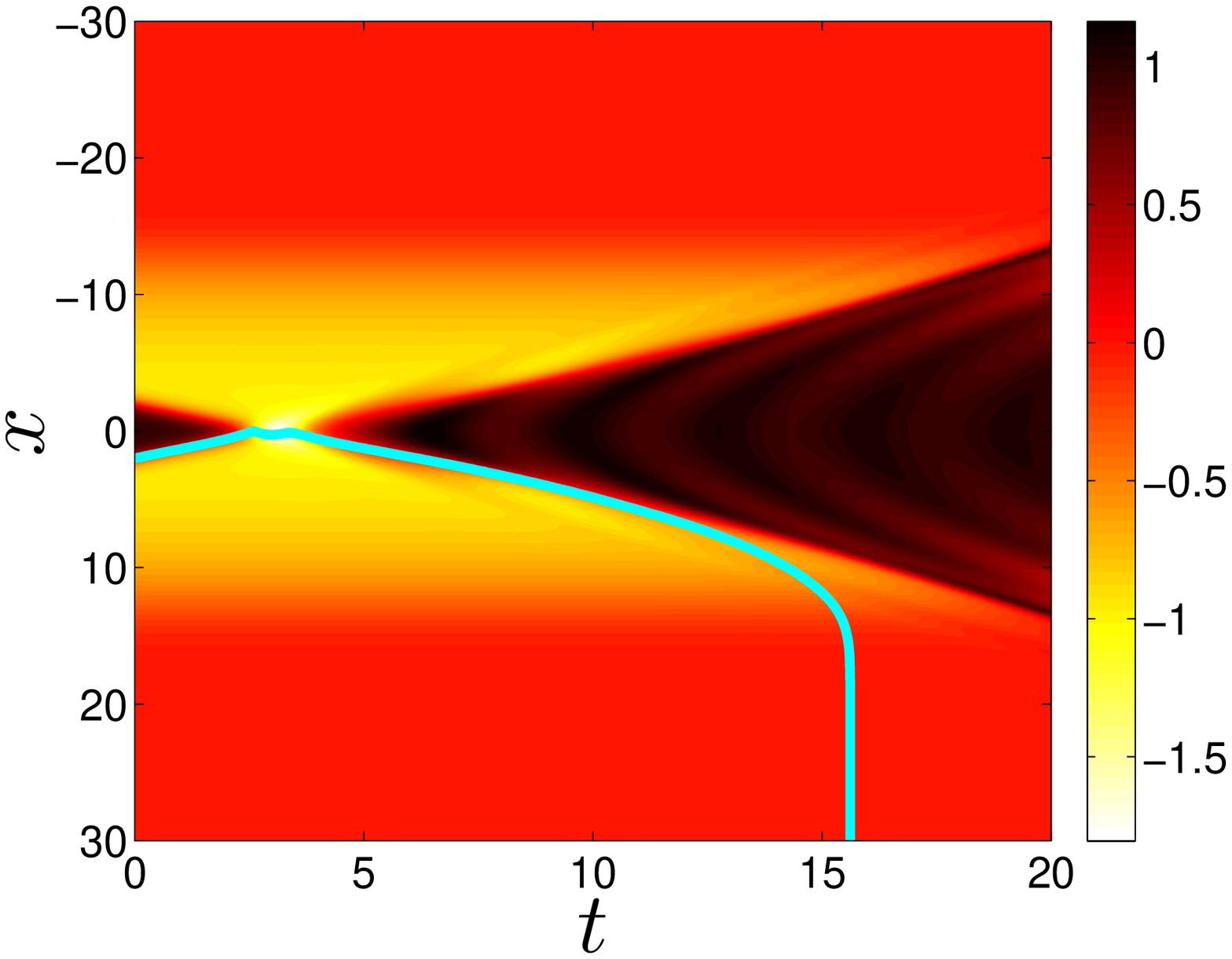}}}
      \subfigure[]{{\includegraphics[width=0.49\textwidth]{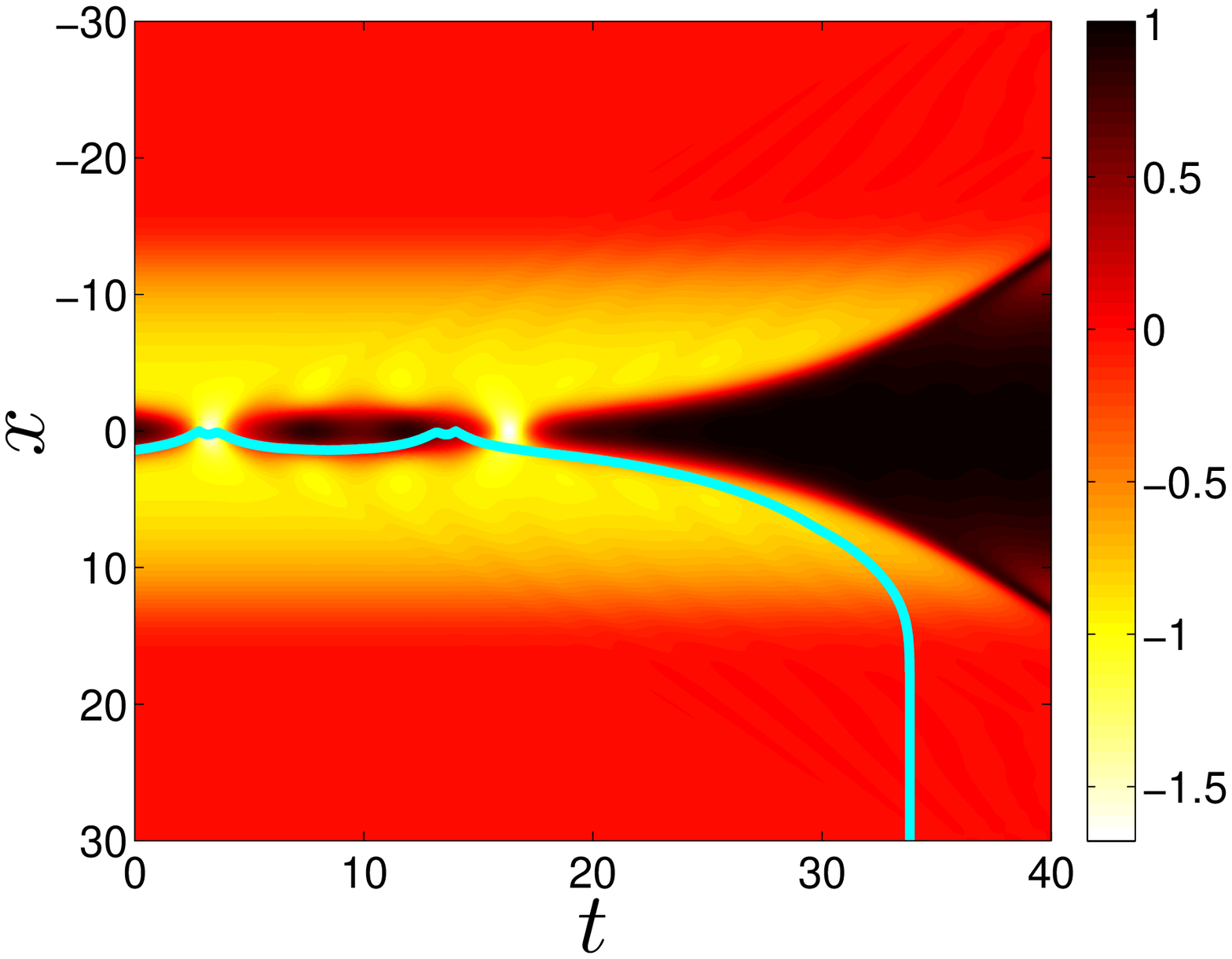}}}
  \end{center}
   \caption{(a) $x_0=2$, $v_{in}=0.7$ (b) $x_0=1.4$, $v_{in}=0.245$ a reflection and two-bounce, for which choosing the same initial speed for the ODE as the PDE yields accurate results.}
    \label{one-two}
\end{figure}

\begin{figure}[tbp]
\begin{center}
      \subfigure[]{{\includegraphics[width=0.49\textwidth]{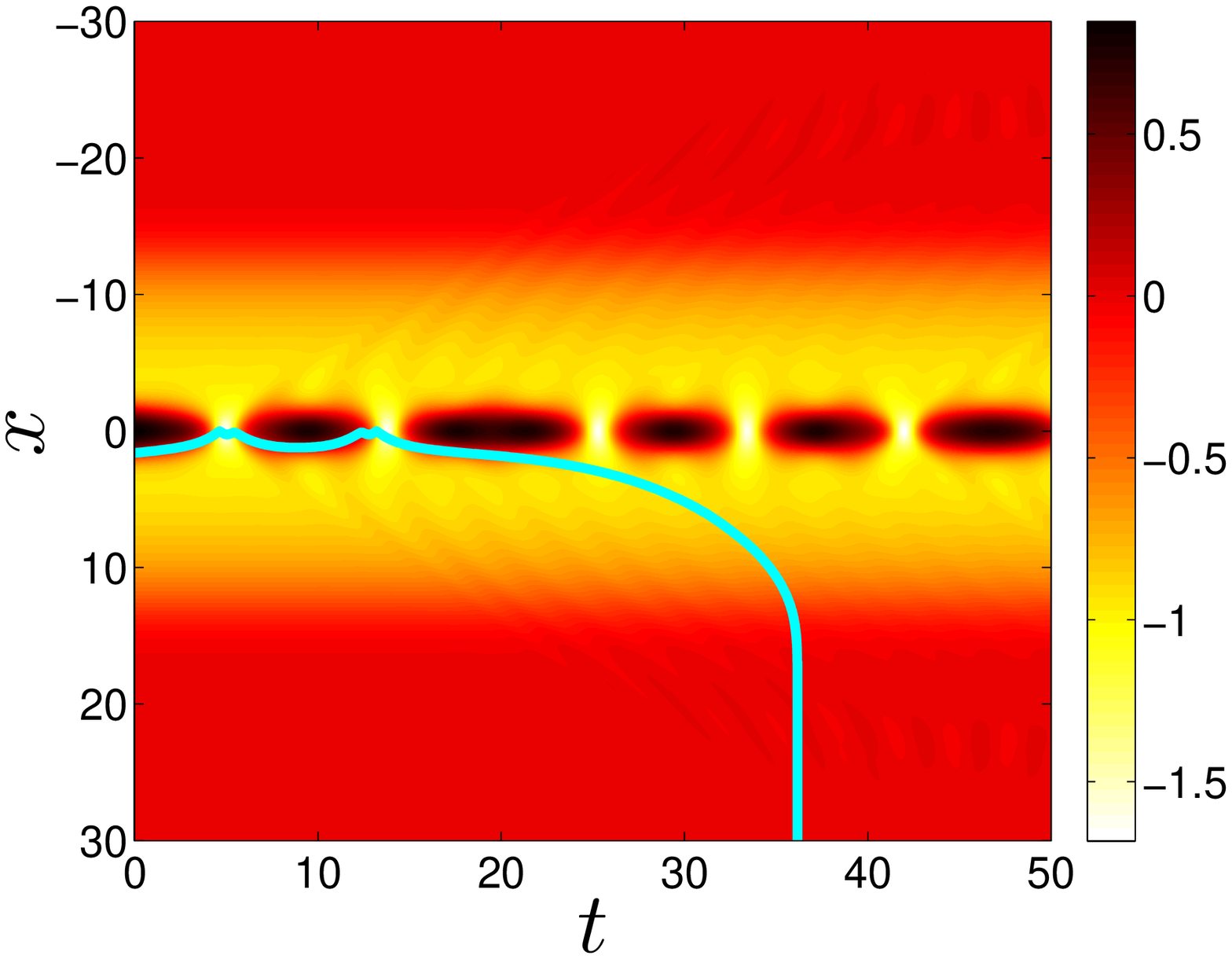}}}
      \subfigure[]{{\includegraphics[width=0.49\textwidth]{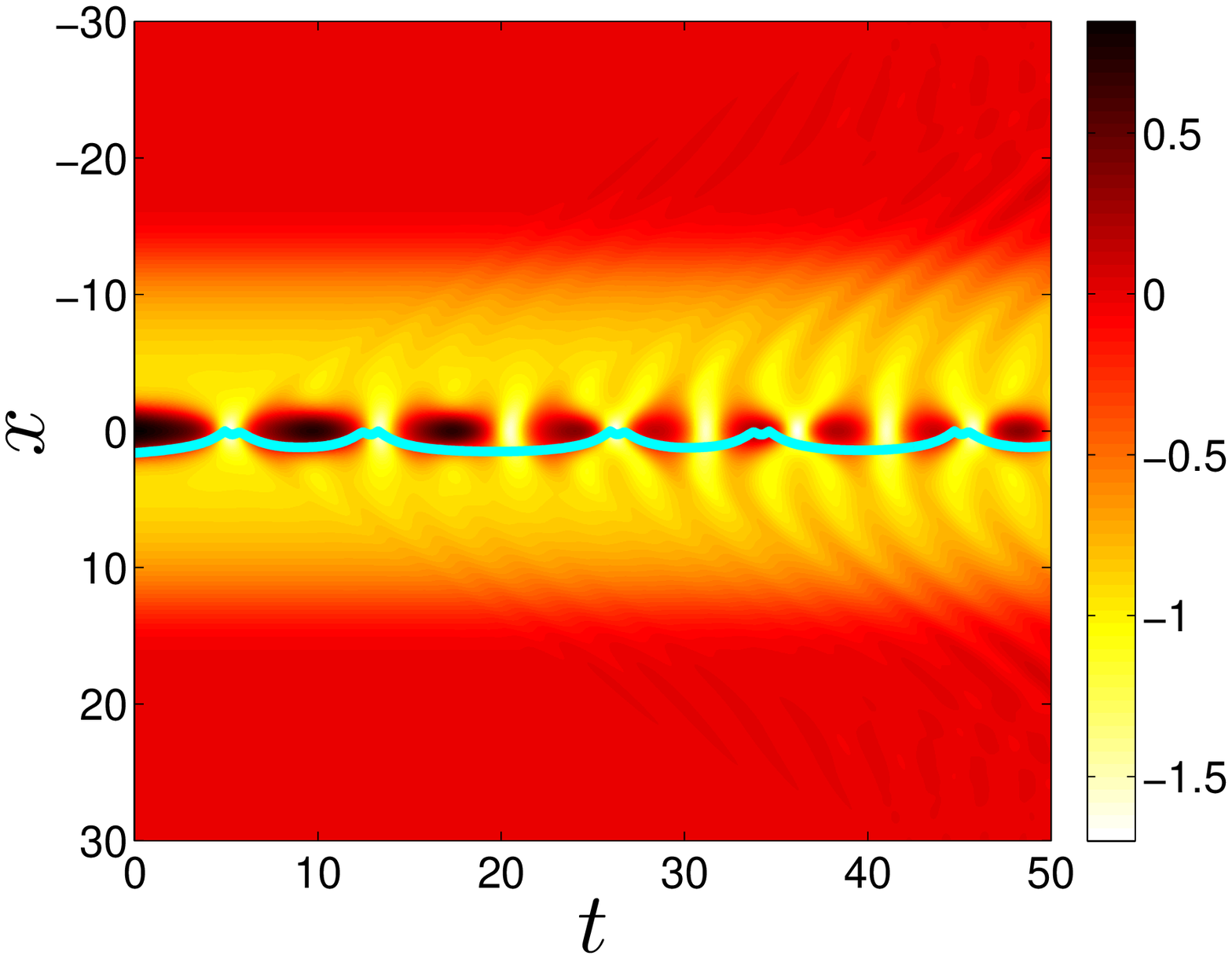}}}
  \end{center}
   \caption{(a) $x_0=1.607$, $v_{in}=0.13$ (b) $x_0=1.607$, $v_{in}=0.11$ a bion, whose formation is not captured very well when the ODE and PDE speeds are the same. In the rightmost figure, the ODE initial speed is decreased by $0.02$. The rightmost ODE solution clearly has better (but still not perfect) agreement with the PDE solution.}
    \label{extra}
\end{figure}

For $\Omega$ values not much larger than the value considered
so far (of $0.15$), the phenomenology is virtually identical to that of the
previous section. All of the interactions we have come across so far,
namely bions, multi-bounce interactions, and expulsion, are observed
for kink-antikink systems as the initial velocity is varied.
Figure \ref{fig:windows_Large_Omega} shows the presence of two- and
three-bounce windows and a threshold reflection velocity for $\Omega=0.2$
and $\Omega=0.3$. However, the widths of existing bounce windows shrink
drastically as the trap strength is increased. Increasing
$\Omega$ from $0.2$ to $0.3$ as in the same figure shrinks the resonance windows by approximately a factor of $\frac{1}{10}$.
Our collective coordinate model also exhibits these features --
it is capable of capturing one- and two-bounce interactions,
and exhibits the same sensitive dependence on the initial speed
we observe when $\Omega$ is small for configurations with more bounces. Examples can be seen in Figure \ref{fig:Omega_02}.

\begin{figure}[tbp]
\begin{center}
      \subfigure[]{{\includegraphics[width=0.49\textwidth]{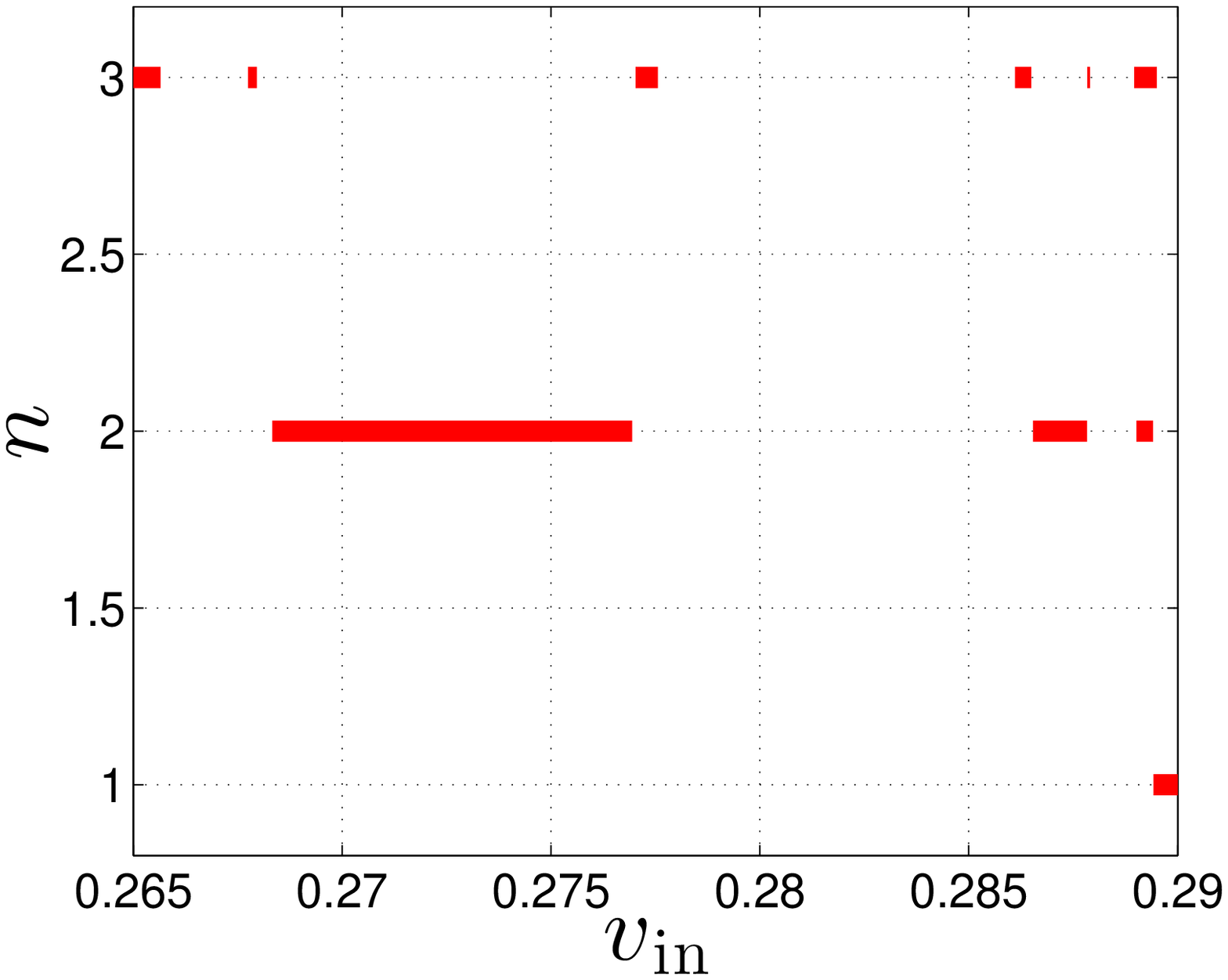}}}
      \subfigure[]{{\includegraphics[width=0.49\textwidth]{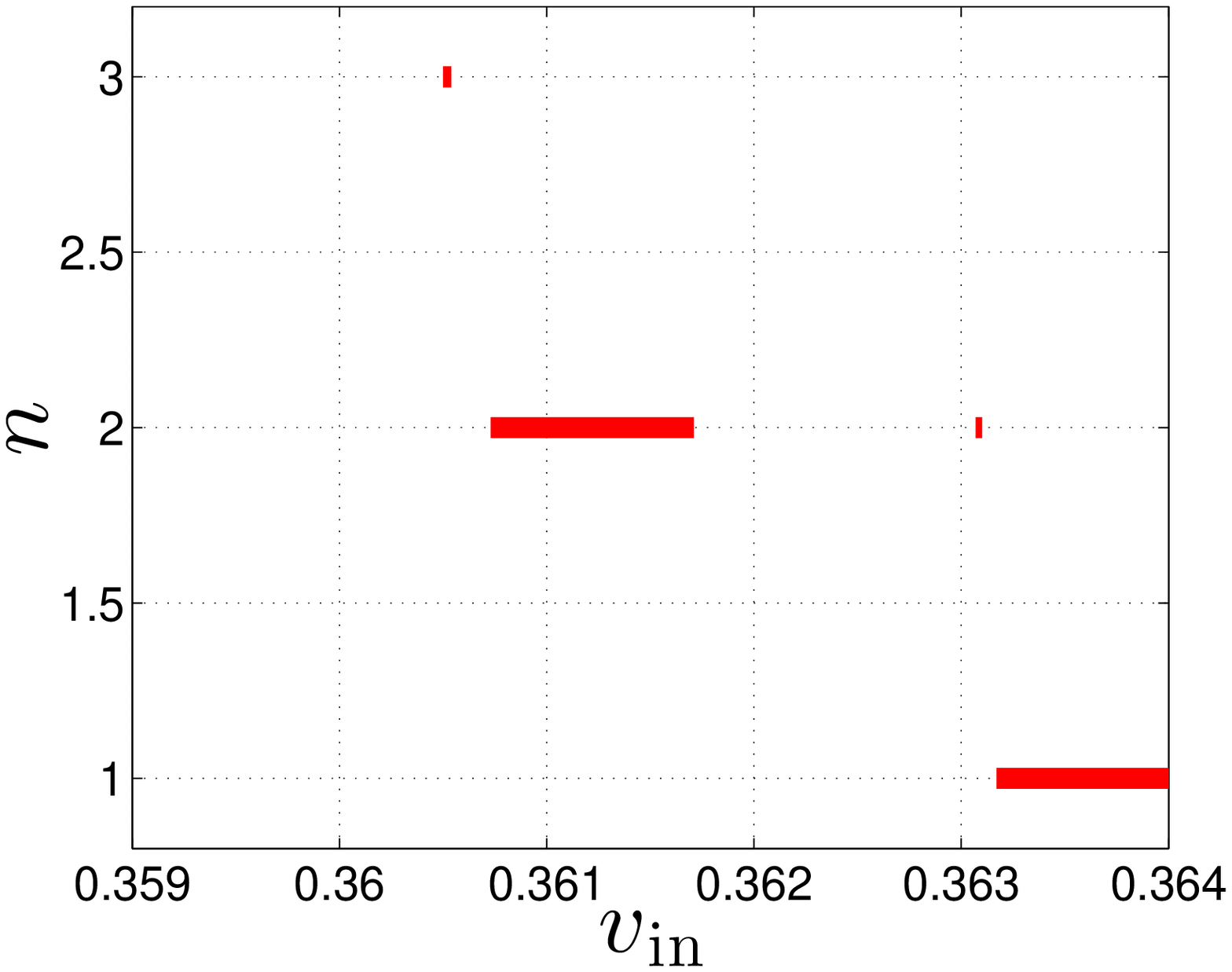}}}
  \end{center}
   \caption{(a) Plots of number of bounces vs. initial velocity for fixed $x_0=2$. (a) $\Omega=0.2$ for $v_\mathrm{in} \in (0.265,0.29)$, (b) $\Omega=0.3$ for $v_\mathrm{in} \in (0.359,0.364)  $.  As $\Omega$ increases, existing multi-bounce windows become significantly narrower (note the change in the velocity increment in the rightmost figure), and many disappear completely.}
    \label{fig:windows_Large_Omega}
\end{figure}

\begin{figure}[h]
	\includegraphics[width=0.32\textwidth]{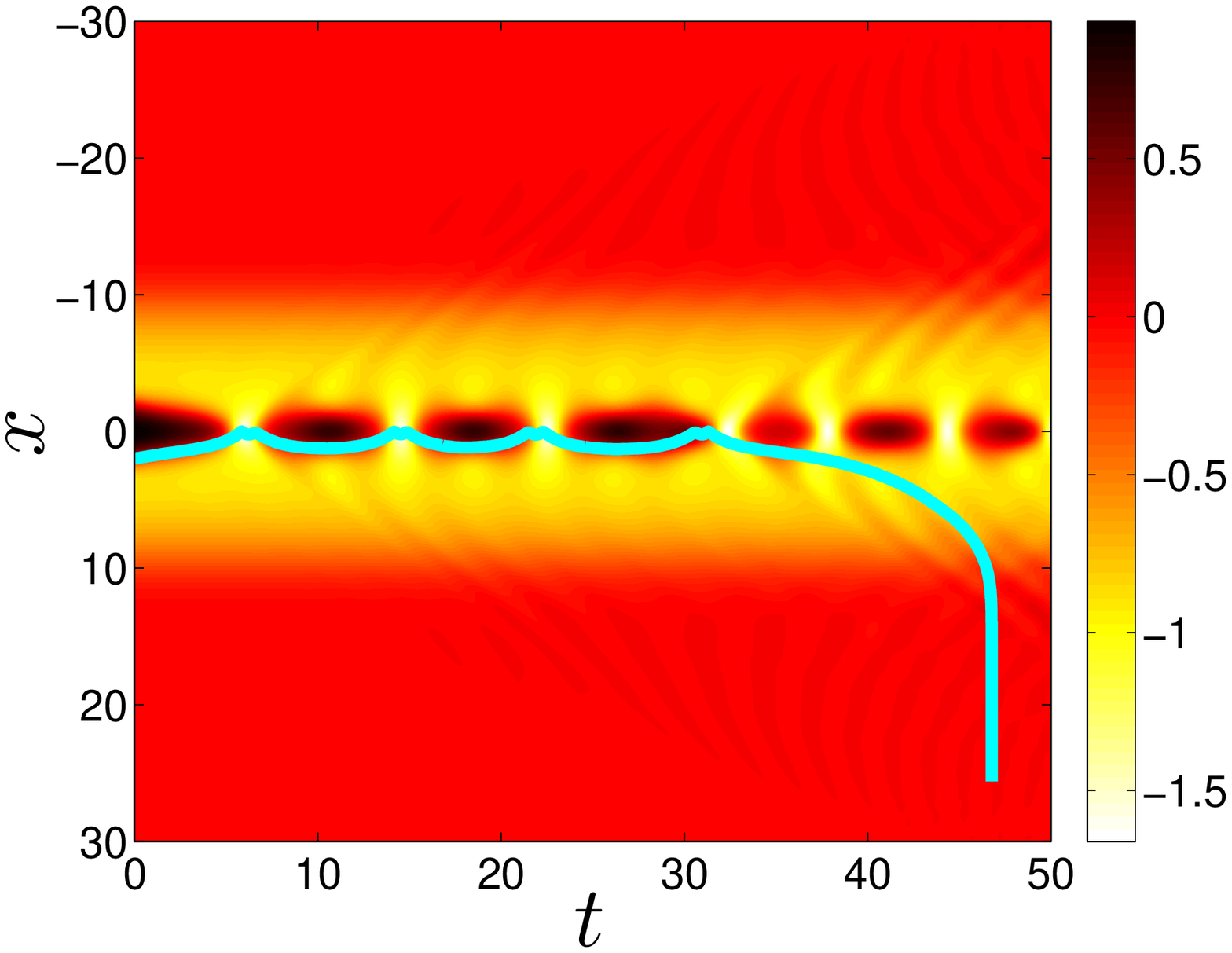}
	\includegraphics[width=0.32\textwidth]{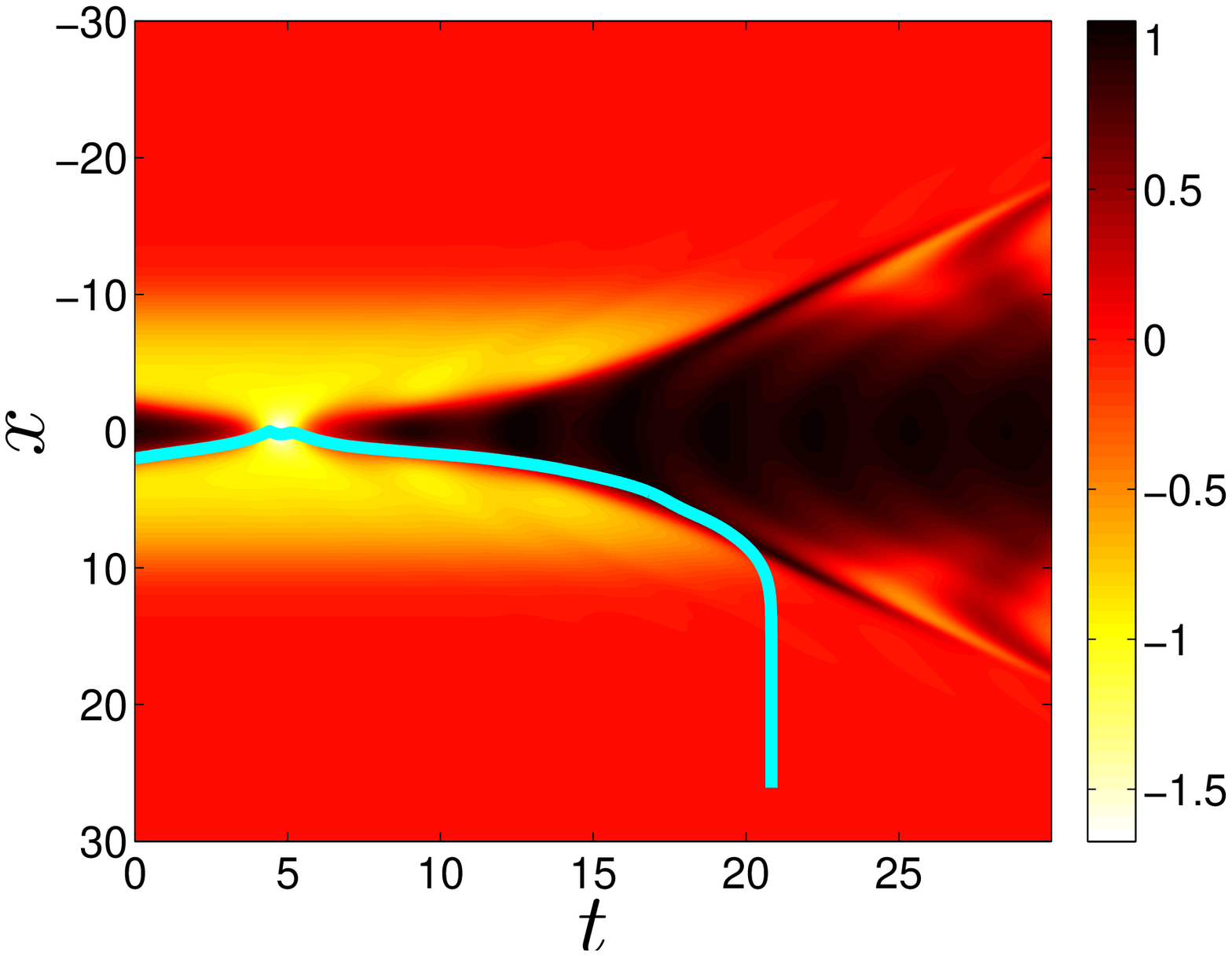}
	\includegraphics[width=0.32\textwidth]{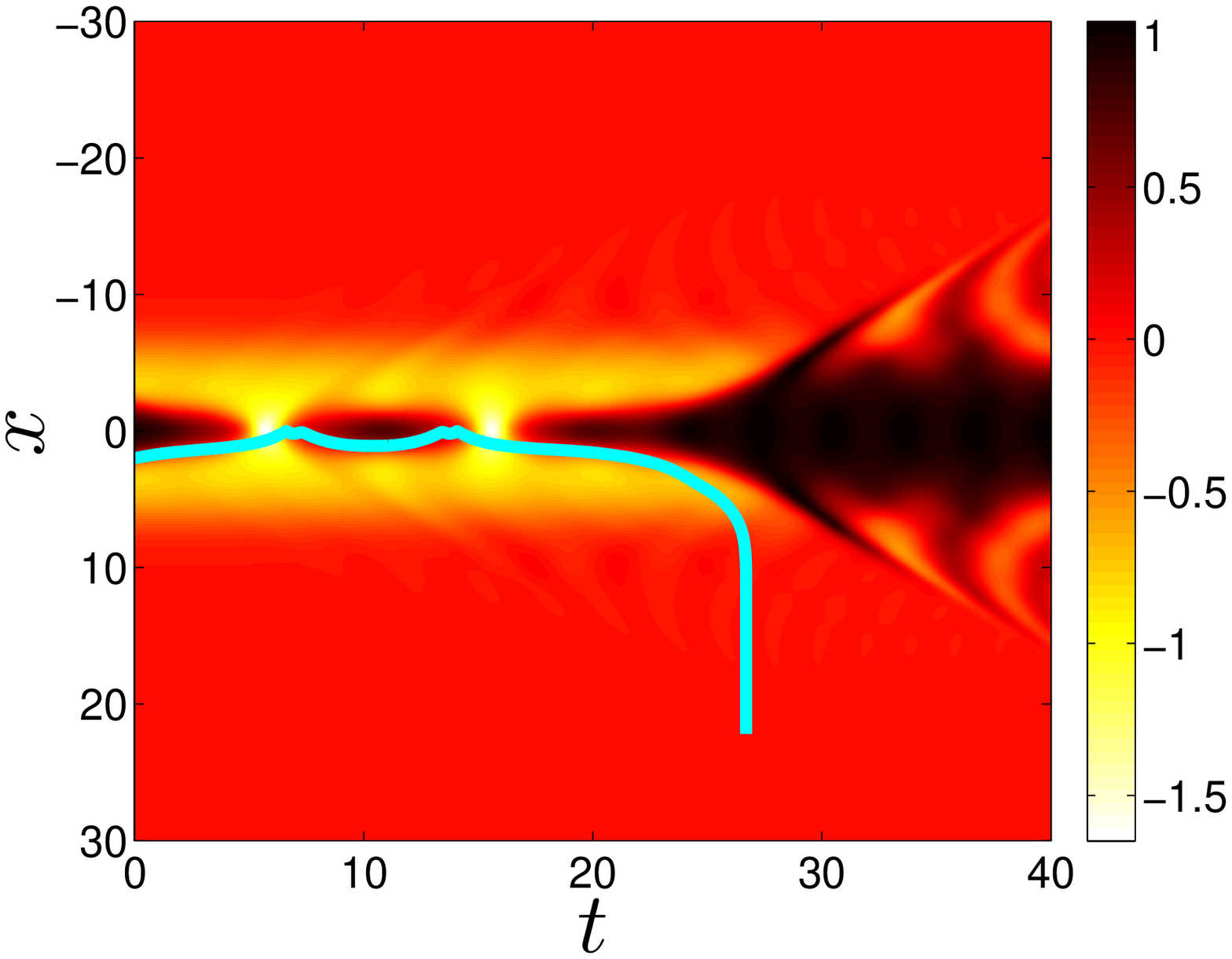}		
	\caption{Collective coordinate results for $\Omega=0.2$ and $\Omega=0.3$ . Here the collective coordinate model performs about as well as when $\Omega=0.15$ in our earlier data runs. $x_0=2$, $v_{in}=0.264$, $\Omega=0.2$ (left), $x_0=2$, $v_{in}=0.364$, $\Omega=0.2$ (middle),  $x_0=2$, $v_{in}=0.361$, $\Omega=0.3$ (right).}
	\label{fig:Omega_02}
\end{figure}

Thus far we have confined ourselves to the case $\Omega \ll 1$. However, the behaviors for larger $\Omega$ are interesting in their own right and can differ significantly from those observed for weak trapping. Intuitively, we expect that a large value of $\Omega$ will constrain the dynamics severely (as
it only allows the kinks to move within a restricted TF region).
Our numerical simulations suggest that two-bounce and three-bounce still exist there, yet
they are extremely narrow and shrink rapidly as $\Omega$ increases.
Figure \ref{big_omega_bounces} shows
some prototypical examples of multi-bounce cases
for $\Omega=0.4$.

\begin{figure}[tbp]
\begin{center}
      \subfigure[]{{\includegraphics[width=0.46\textwidth]{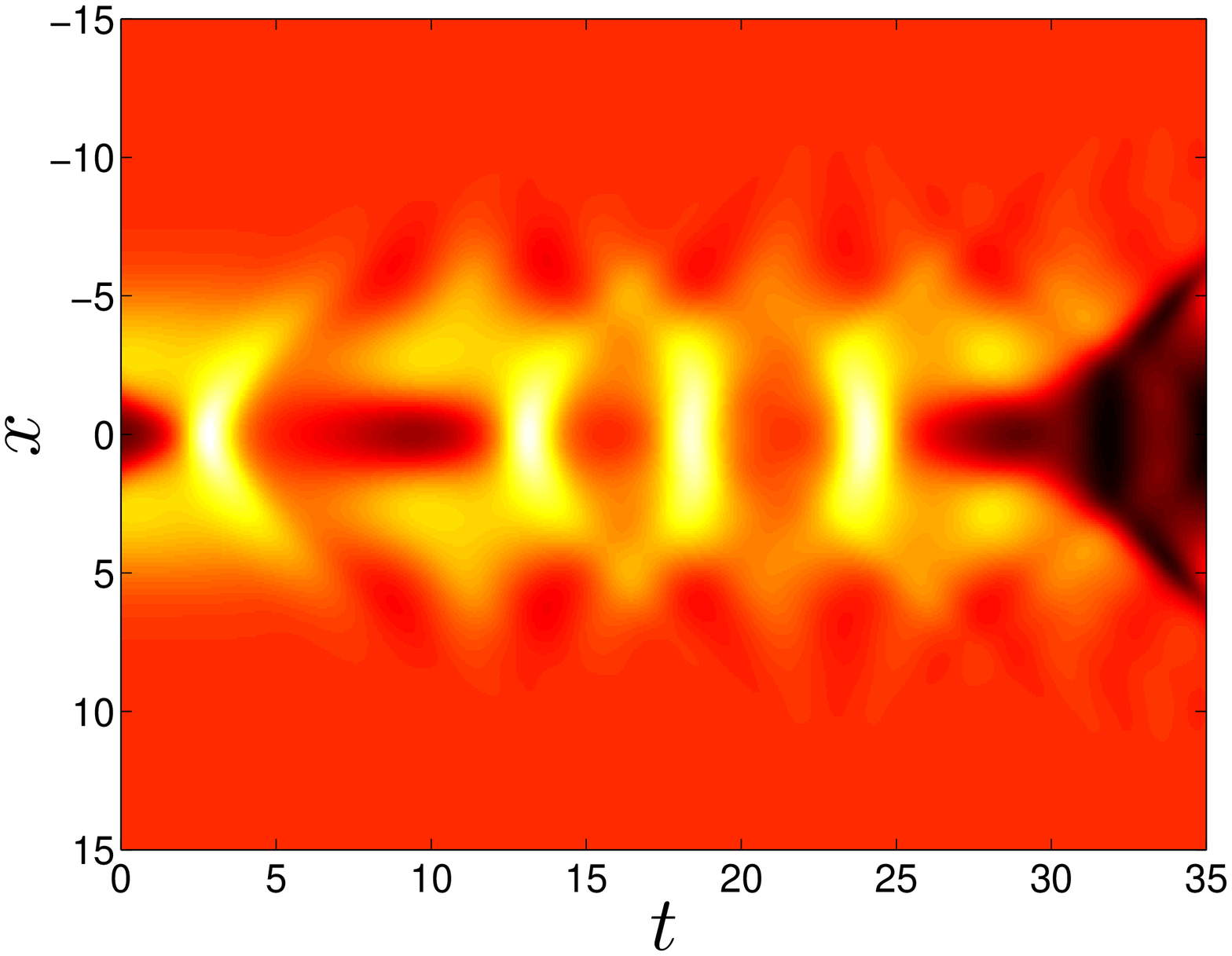}}}
      \subfigure[]{{\includegraphics[width=0.49\textwidth]{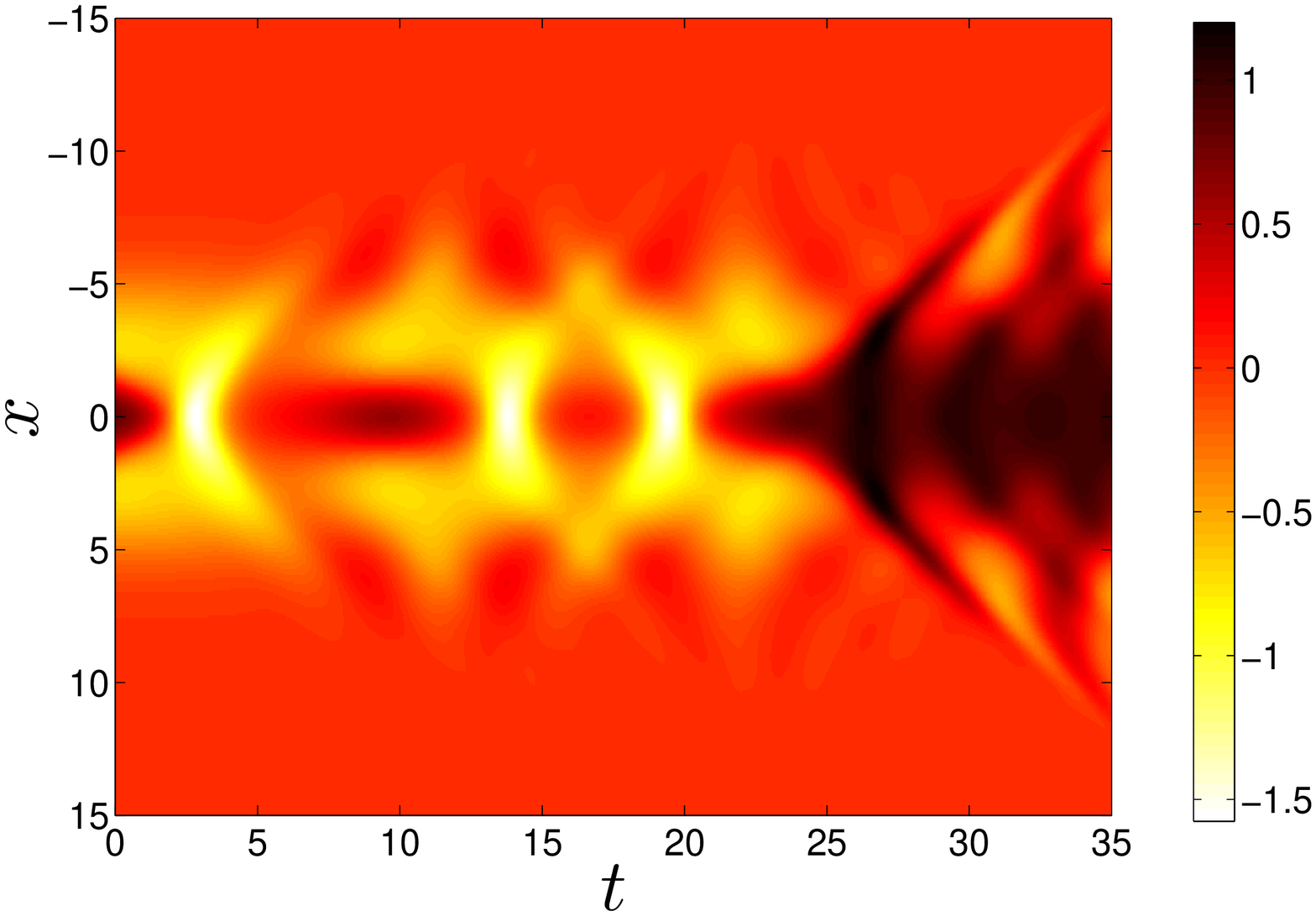}}}
  \end{center}
    \caption{ Kink-antikink system for $\Omega=0.4$, $x_0=1.4$, (a) $v_{in}=0.446$ (b) $v_{in}=0.4463$. }
\label{big_omega_bounces}
\end{figure}

\section{Conclusions and Future Challenges}
In this work we have taken the well-known $\phi^4$ model in a
novel direction: motivated by considerations in atomic BECs
where a complex variant of the model is examined in the presence
of an external potential, we have explored how the standard
$\phi^4$ phenomenology is affected by a parabolic trapping.
We have found that the results can be fundamentally different
in comparison to the standard homogeneous (untrapped) case.
In particular, even the single kink state is dynamically
unstable, representing a saddle point in configuration
space and leading to the expulsion of the kink from the
trap center of symmetry. Moreover, in the presence of
multiple kinks and anti-kinks, the number of coherent structures
determines the number of unstable eigendirections (and the
dimension of the corresponding unstable manifold).
It is interesting to also observe the reverse role of
the potential and of the kink-antikink interaction in comparison
with the atomic BEC case. These features generally favor
the expulsion of the kinks from the system unless the kinks
are sufficiently close to each other that their interaction
dominates. In that setting, $n$-bounce windows may still be present,
although these windows were found to shrink radically
as the potential parameter $\Omega$ is increased.
Lastly, we have developed a collective coordinate low-dimensional
ODE approach, under relevant approximations (such as the truncation
of the higher power terms) which yields, especially for low
numbers of bounces, reasonably consonant results between the
PDE field theory and the ODEs. On the other hand, when presenting
a more ``complete'' variant of the equations, issues similar
to those encountered by~\cite{weigel,weigel2} in the homogeneous
variant of the model emerge. 
Indeed, the prospect of resolving these collective coordinate issues and improving the
ODE models in this inhomogeneous $\phi^4$ variant (and even
for the homogeneous case within the model) is an important
avenue for possible future work.

From a mathematical standpoint, understanding better the
stability properties of the model (e.g. proving rigorously
the eigenvalue estimate in the case of $N$ kinks, as likely
possessing $N$ unstable eigendirections) could be
an interesting direction for further work. Also, extending
inhomogeneous model considerations to other models of
recent interest such as the $\phi^6$~\cite{shnir1,weigel,gani1,danial,usrecent},
and $\phi^8$ models~\cite{gani4,gani5,ivanrecent} would
be valuable in understanding the genericity
(or not) of the conclusions drawn in this study. Some of these
directions are currently under consideration and will
be reported in future publications.

\appendix
\section{Tables for $n-$ bounce windows}
We list the velocity intervals $[v_1,v_2]$ that result in multi-bounce windows when the initial velocity $v_\mathrm{in}$ is picked within these intervals.  In Tables 1--2,  we list those intervals for a small separation ($x_0=1.4$) and a big separation ($x_0=7$) with $\Omega=0.15$. In Tables 3--4, we list those intervals for bigger values of $\Omega$ ($0.2$ and $0.3$ respectively) with fixed $x_0=2$. 

\begin{table}\label{mb-x0_14}
\begin{center}
{
 \begin{tabular}{|l|c|c|c|}
		\hline
		$n$ & $v_1$ & $v_2$ & $\Delta v_n$ \\
		\hline
		3 & 0.23729 & 0.23827 & 0.00098  \\
		
		3 & 0.23973 &  0.23996 & 0.00023 \\
		3 & 0.24023& 0.24035 & 0.00012 \\

		2 & 0.24038 & 0.24744 & 0.00706 \\
		3 & 0.24786 &0.24801 & 0.00015 \\
		3 & 0.25396 & 0.2545& 0.00054 \\
		2 & 0.25453 & 0.25601 & 0.00148 \\ 
		2 & 0.25754 & 0.25787 & 0.00033 \\
		2 & 0.25824 &0.25831 & 0.00007 \\
		2 & 0.2584 & 0.25841& 0.00001 \\

		\hline
	\end{tabular}

\caption{n-bounce windows for $x_0=1.4$. One-bounce windows occur for $v_\mathrm{in}>0.25845$.}
}
\end{center}
\end{table}

\begin{table}\label{mb-x0_7}
\begin{center}
{
\begin{tabular}{|l|c|c|c|}
		\hline
		$n$ & $v_1$ & $v_2$ & $\Delta v_n$ \\
		\hline
		2 & 0.73116 &  0.7314 & 0.00024 \\

		2 & 0.73260 & 0.7331 & 0.00050 \\
		3 & 0.73327 & 0.73331 &  0.00004 \\
		
		2 & 0.73527 & 0.73686 & 0.00159 \\
		 3& 0.73791 &0.73981 & 0.0019\\
		 
		 
		 2& 0.74083 & 0.74214& 0.00131\\
		 
		 3&0.74226  &0.74228  & 0.00002 \\
		 
		 3&0.74346  & 0.74353& 0.00007 \\
		 
		 2&0.74355  &0.74376& 0.00021 \\
		 2&0.74401  & 0.74411 & 0.0001\\

		\hline
		\end{tabular}

\caption{n-bounce windows for $x_0=7$. One-bounce windows occur for $v_\mathrm{in}>0.74414$.}

}
\end{center}
\end{table}

\begin{table}\label{mb-x0_2_1}
\begin{center}
{
 \begin{tabular}{|l|c|c|c|}
		\hline
		$n$ & $v_1$ & $v_2$ & $\Delta v_n$ \\
		\hline
		3 & 0.26501 &  0.26565 & 0.00064  \\
		3 & 0.26775 & 0.26796 & 0.00021 \\
		2 & 0.26833 & 0.27694 & 0.00861 \\
		3 & 0.27703 &0.27756 & 0.00053 \\
		3 &0.28611  & 0.28649& 0.00038 \\
		2 & 0.28654 & 0.28783 & 0.00129 \\ 
		3 & 0.28784 &0.2879 & 0.00006\\
		3 & 0.28896 & 0.28897 & 0.00001\\
		2 &0.28902 &  0.28941& 0.00039 \\

		\hline
	\end{tabular}

\caption{n-bounce windows for $x_0=2$, $\Omega=0.2$. One-bounce windows occur for $v_\mathrm{in}>0.28942$.}
}
\end{center}
\end{table}

\begin{table}\label{mb-x0_2_2}
\begin{center}
{
 \begin{tabular}{|l|c|c|c|}
		\hline
		$n$ & $v_1$ & $v_2$ & $\Delta v_n$ \\
		\hline
		3 & 0.3605 &  0.36054 & 0.00004  \\
		2 & 0.36073 & 0.36171 & 0.00098 \\
		2 & 0.36307& 0.3631 & 0.00003\\
						
		\hline
	\end{tabular}

\caption{n-bounce windows for $x_0=2$, $\Omega=0.3$. One- bounce windows occur for $v_\mathrm{in}>0.36317$.}
}
\end{center}
\end{table}

\newpage
\section{Derivation of the coefficients in Eq. (\ref{lag1})}
We define $\phi_{\pm}=\pm \phi_0(x\pm X(t))$ and $\chi_{\pm}=\pm \chi_{1}(x\pm X(t))$, $\phi'_{\pm}=\pm \phi_0'(x\pm X(t))$ and $\chi'_{\pm}=\pm \chi'_1(x\pm X(t))$. Then Eq.~(\ref{ansatz}) becomes
\begin{equation} \label{kink-antikink_2}
u(x,t)=u_{\Omega}(\phi _+ + \phi_{-} -1) + A(\chi_{+}+\chi_{-}).
\end{equation}

Substituting (\ref{kink-antikink_2}) into Eq.~(\ref{lagr}) gives 
\begin{equation}
\begin{aligned}
 L=&\int  \left\{\frac{1}{2}\left[ u_{\Omega}\left( \phi _{+}^{\prime
}-\phi _{-}^{\prime }\right) \dot{X}  +\dot{A}(\chi_{+}+\chi_{-})+A(\chi_{+}^{\prime }-\chi_{-}^{\prime
})\dot{X}
 \right]^{2} \right\}\mathrm{d}x\\
 -&\int  \left\{\frac{1}{2}\left[ u_{\Omega}\left( \phi _{+}^{\prime
}+\phi _{-}^{\prime }\right)   +u_{\Omega}'(\phi _+ + \phi_{-} -1) +A(\chi_{+} ^{\prime }+\chi_{-}^{\prime })
\right]^{2} \right\}\mathrm{d}x\\
-&\int V(u) \mathrm{d}x.
\label{Lagrangian_expanded}
\end{aligned}
\end{equation}
To handle the $V(u)$ terms we first write $u =u _{a}+u _{b}$ where $u
_{a}=u_{\Omega}(\phi _+ + \phi_{-} -1)
$ and $u _{b}= A(\chi_{+}+\chi_{-})$. Then using the Taylor series expansion, we get
\begin{equation}
\begin{aligned}
V(u )&= V(u _{a}+u _{b})\\
&=V(u _{a})+V^{\prime }(u
_{a})u _{b}+\frac{V^{\prime \prime }(u _{a})}{2!}u _{b}^{2}+\frac{V^{^{\prime \prime \prime }}(u _{a})}{3!}u _{b}^{3}+\frac{V^{(4)}(u
	_{a})}{4!}u _{b}^{4} 
\label{eq:taylor_V}
\end{aligned}
\end{equation}

The corresponding reduced Lagrangian (ignoring higher order terms) that is used in our simulations is given by Eq. (\ref{lag1}). Applying Euler-Lagrange equations, we obtain Eq. (\ref{EL_ode}).

We will list the formulae of the coefficients below. Note that they are all functions of $X(t)$. Since $\chi_{1}$ is not known explicitly, the coefficients presented are in the integral form. These coefficients are calculated numerically.

\begin{eqnarray*}
U(X)=&\int \frac{1}{2}\left( u_{\Omega} \left( \phi _{+}^{\prime
}+\phi _{-}^{\prime }\right) +u_{\Omega}'(\phi _{+}+\phi _{-}-1)\right) ^{2}\,\mathrm{d}x
\\
&+\int V(u_{\Omega}(\phi _{+}+\phi _{-}-1))\,\mathrm{d}x
\end{eqnarray*}
\begin{eqnarray*}
\begin{aligned}
F(X)=-\frac{1}{2}\int &(u_{\Omega} \left( \phi _{+}^{\prime
}+\phi _{-}^{\prime }\right) +u_{\Omega}'(\phi _{+}+\phi _{-}-1))(\chi _{+}^{\prime }+\chi
_{-}^{\prime })\,\mathrm{d}x\\
 &-\frac{1}{2}\int V^{\prime }(u_{\Omega}(\phi _{+}+\phi _{-}-1))(\chi
_{+}+\chi _{-})\,\mathrm{d}x
\end{aligned}
\end{eqnarray*}
\begin{eqnarray*}
\begin{aligned}
K(X)=&-\frac{1}{2}\int(\chi _{+}^{\prime }+\chi _{-}^{\prime
})^{2}\,\mathrm{d}x
-\frac{1}{2}\int V^{\prime \prime }(u_{\Omega}(\phi _{+}+\phi
_{-}-1))(\chi _{+}+\chi _{-})^{2}\,\mathrm{d}x
\end{aligned}
\end{eqnarray*}
\begin{eqnarray*}
I(X)=\frac{1}{2}\int u_{\Omega}^2\left( \phi _{+}^{\prime }-\phi
_{-}^{\prime }\right) ^{2}\,\mathrm{d}x, \hspace{0.5cm}Q(X)=\frac{1}{2}\int (\chi _{+}+\chi _{-})^{2}\,\mathrm{d}x, 
\end{eqnarray*}

\begin{eqnarray*}
C(X)=\frac{1}{2}\int u_{\Omega}\left( \phi _{+}^{\prime }-\phi _{-}^{\prime }\right) (\chi
_{+}+\chi _{-})\,\mathrm{d}x
\end{eqnarray*}


\printindex
\end{document}